%% file: main.tex
\documentclass[sigconf,10pt]{acmart}
\setcopyright{none}
\renewcommand\footnotetextcopyrightpermission[1]{}
\acmConference{}{}{}
\acmBooktitle{}
\acmYear{}
\acmISBN{}
\acmDOI{}

\usepackage{ifthen}
\usepackage{amsmath}
\usepackage{xspace}
\usepackage{graphicx}
\usepackage{booktabs}
\usepackage{subcaption}
\usepackage{enumitem}
\newboolean{showcomments}
\setboolean{showcomments}{true}
\ifthenelse{\boolean{showcomments}}
{ \newcommand{\mynote}[3]{
    \protect\fbox{\bfseries\sffamily\scriptsize#1}
    {\small$\blacktriangleright$\textsf{\emph{\color{#3}{#2}}}$\blacktriangleleft$}}}
{ \newcommand{\mynote}[3]{}}

\newcommand{\todo}[1]{{\color{orange}{\textbf{\em [TODO: #1]}}}\xspace}

\AtBeginDocument{%
  }

\author{Muaz Ali}
\authornote{These authors contributed equally to this paper.}
\affiliation{\institution{University of Arizona} \country{Tucson, AZ, USA}}

\author{Utkarsh Upadhyay}
\authornotemark[1]
\affiliation{\institution{University of Arizona} \country{Tucson, AZ, USA}}

\author{Sean McCormick}
\affiliation{\institution{University of Arizona} \country{Tucson, AZ, USA}}

\author{Joseph Hill}
\affiliation{\institution{University of Arizona} \country{Tucson, AZ, USA}}

\author{Beichuan Zhang}
\affiliation{\institution{University of Arizona} \country{Tucson, AZ, USA}}

\renewcommand{\shortauthors}{Ali and Upadhyay}

\begin{document}

\title{Starlink Constellation: Deployment, Configuration, and Dynamics}

\begin{abstract}

Starlink has rapidly emerged as the world's largest satellite constellation and the \textit{de facto} reference system for low Earth orbit (LEO) networking research. Existing literature predominantly models Starlink as a static, symmetric, and fully deployed structure with uniformly distributed satellites. However, we reveal that Starlink's actual deployment, orbital configurations, and operational dynamics fundamentally deviate from these idealized assumptions. 

Leveraging satellite observation data spanning 2019 to 2025, we demonstrate that the constellation is highly dynamic across multiple temporal and spatial scales. Macroscopically, Starlink comprises multiple orbital shells undergoing continuous active deployment and reconfiguration. Microscopically, individual satellites exhibit high mobility, frequently executing collision-avoidance maneuvers, altitude adjustments, and intra-orbital relocations. We discover that while the majority of satellites form a relatively stable structure with near-uniform spacing, other satellites tend to cluster as twins or triads as in-orbit backups. Furthermore, empirical survival analysis indicates an operational lifespan of 4--6 years and an average daily failure probability of $0.0128$\%.

Ultimately, our data-driven characterization exposes Starlink as a highly heterogeneous and continuously evolving network. We provide critical empirical insights that challenge prevailing simulation models, offering a more accurate foundation for future LEO topology design, routing protocols, and performance evaluations.

\end{abstract}

\maketitle


\input{introduction}

\input{background}
\input{analysis}

\input{implications_network_design}
\input{related_works}

\input{conclusion}

\bibliographystyle{ACM-Reference-Format}
\bibliography{references}
\input{appendix}

\end{document}

%% file: introduction.tex
\section{Introduction}

Massive Low Earth Orbit (LEO) satellite constellations have transformed global broadband connectivity. At the forefront is SpaceX's Starlink, which, with more than 9,000 satellites by the end of 2025, is the world's largest commercial constellation. It has solidified its position as the \textit{de facto} reference architecture for the academic networking community, relied upon heavily to design and evaluate next-generation LEO topologies, routing protocols, and traffic engineering strategies~\cite{sate,motif,3_isl_grid}.

Despite this, current research predominantly models Starlink using pristine, static Walker configurations derived from FCC filings. Literature routinely assumes symmetric, fully deployed structures with perfectly uniform spacing and deterministic trajectories~\cite{motif,mcts,hetrogenous_topology,delyHandley,sate,routing_1,3_isl_grid,IPCGrid,hypatia,deepRL_routing}. Protocols are evaluated under the premise that the topology, once established, remains continuously stable.

However, this static view fails to account for the severe challenges of operating such a large-scale system in space, and to date, there has been no systematic study to characterize the actual in-orbit reality of these mega-constellations. In this paper, we conduct the first comprehensive longitudinal study into this reality using six years of satellite observation data (2019--2025). We reveal that the reality of operating in LEO fundamentally contradicts static assumptions. Driven by harsh environmental constraints and continuous operational requirements, the network is in a perpetual state of flux. 
Macroscopically, the constellation is composed of multiple orbital shells, most of which are under active deployment and reconfiguration at any given time. We have not seen any major shell that is fully deployed and stable over time. Microscopically, individual satellites frequently alter trajectories to avoid potential collisions with other space objects, and adjust altitudes or intra-orbit positions for operational purposes. 


Interestingly, we find that the intra-orbit satellite spacing is not uniform as assumed by prior work. The majority of the satellites on the same orbital plane are placed with a fixed spacing or multiples of it, which allows them to form a relatively stable backbone structure. We term these satellites \textit{regular} satellites. However, there are also a significant number of \textit{non-regular} satellites that deviate from this structure, staying close to regular satellites to form \textit{twins} or \textit{triads}, likely as hot standbys to provide redundancy in case of failures. 
Finally, our empirical survival analysis reveals that satellites have an operational lifespan of 4 to 6 years and an average daily failure probability of $0.0128$\%.

To demonstrate the impacts of these findings, we conduct a case study comparing the grid topology's network performance under perfect constellation versus the real-world constellation. We show that the dynamic nature of the constellation leads to significant route flapping and latency spikes when using existing protocols designed for static topologies. By exposing Starlink as a heterogeneous and evolving network, our findings provide the empirical foundation necessary to design robust LEO topologies and routing protocols that reflect the chaotic reality of space.


In summary, this paper makes the following key contributions:
\begin{itemize}
    \item \textbf{Longitudinal Empirical Analysis:} We present a data-driven characterization of Starlink, exposing discrepancies between idealized models and in-orbit reality.
    \item \textbf{Multi-Scale Dynamics:} We map macroscopic shell evolution and quantify microscopic mobility, including collision-avoidance, altitude adjustments, and intra-orbit re-positioning.
    \item \textbf{Intra-Orbit Configurations:} We identify the operational topological behavior of 'regular' versus 'non-regular' (twin/triad) satellite clustering.
    \item \textbf{Lifecycle Analysis:} We provide an empirical survival analysis detailing operational lifespans and baseline failure probabilities.
\end{itemize}

The rest of the paper is organized as follows. Section~\ref{sec:background} presents background on LEO constellations, prior modeling assumptions, and our data sources. Section~\ref{sec:analysis} introduces our research questions and presents empirical findings on deployment, configuration, and constellation dynamics. Section~\ref{sec:implications_network_design} discusses the implications of these findings for network design. Section~\ref{sec:related_works} reviews related work, and Section~\ref{sec:conclusion} concludes.

%% file: background.tex
\section{Constellation Model and Data Pipeline}
\label{sec:background}

We first introduce the orbital concepts used in our analysis, then describe the idealized constellation structure specified in Starlink's deployment filings, how it is commonly used in prior models, and the data used in our study.

\subsection{Orbital Parameters}
\begin{figure}[t]
    \centering
    \includegraphics[width=0.7\linewidth]{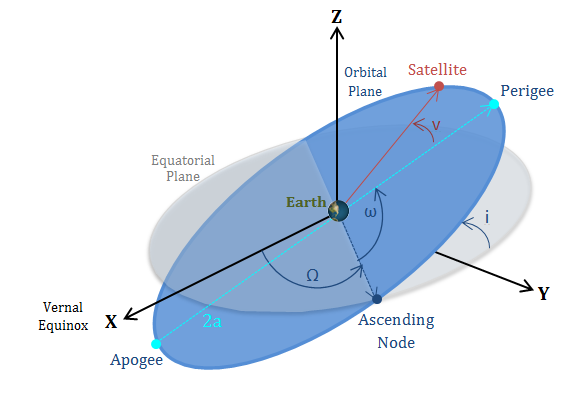}
    \caption{Orbital parameters of a LEO Satellite~\cite{orbital_parameters}.}
    \label{fig:orbital_technical_parameters}
\end{figure}

A LEO constellation contains thousands of satellites organized into multiple shells, each defined by a common inclination and altitude. Altitude is the distance above Earth's surface, while inclination is the tilt of the orbit relative to Earth's equator. LEO satellites operate at altitudes below 2{,}000\,km, and a satellite typically orbits the Earth in around two hours. Within each shell, satellites are distributed across many orbital planes. Satellites in the same orbital plane follow nearly the same path around Earth and therefore share the same, or very similar, values of the right ascension of the ascending node (RAAN). As shown in Figure~\ref{fig:orbital_technical_parameters}, the RAAN, denoted by $\Omega$, is the angle in the equatorial plane that specifies the orientation of an orbital plane around Earth. To describe a satellite's position within that plane, we use its phase, denoted by $u$. For the near-circular LEO orbits considered in this paper, we compute $u$ as the argument of latitude, defined as the sum of the argument of perigee $\omega$ and the true anomaly $\nu$, i.e., $u=\omega+\nu$, both illustrated in Figure~\ref{fig:orbital_technical_parameters}. The angle $u$ is measured from the ascending node to the satellite's current position along the orbital plane. This phasing angle therefore provides a consistent way to compare the relative positions of satellites within the same orbital plane, since all positions are measured from the same reference point.

\subsection{Real  and  Perfect Constellation}

Starlink's FCC filings with the Federal Communications Commission specify the intended structure of each shell, including its altitude, inclination, number of orbital planes, and number of satellites per plane~\cite{fcc_eratum23,fcc2021starlink}. These filings provide the design blueprint for the constellation and are widely used as baseline configurations in prior LEO networking studies. A commonly used reference shell is Starlink's first shell, which consists of 72 orbital planes with 22 satellites per plane. Most prior studies model LEO constellations using the Walker-delta design, denoted by $i$:$T$/$P$/$F$, where $i$ is the inclination, $T$ is the total number of satellites, $P$ is the number of orbital planes, and $F$ is the relative phase offset between adjacent planes. This model assumes a fixed number of planes, a uniform number of satellites per plane, and uniform phase spacing. Under this abstraction, shells are treated as fully deployed and symmetric, typically by mapping parameters from regulatory filings directly to a Walker-delta configuration.


In practice, however, the deployed constellation does not always follow this idealized structure. Satellites are launched gradually, maneuvered during operation, and eventually deorbited, while real-world factors such as solar storms and collision-avoidance maneuvers introduce further dynamics. The constellation deployment plan also changes~\cite{fcc_eratum23}. These effects make shells appear irregular at different stages of deployment and create a gap between the planned constellation and the system that actually exists in orbit. Because network behavior depends on the realized orbital structure, understanding Starlink requires analyzing the deployed system rather than relying only on its idealized blueprint. In the rest of this paper, we quantify how the real constellation departs from the ideal design and how its structure evolves over time.

\subsection{Dataset and Analysis Pipeline}
In contrast to traditional terrestrial networks, a unique advantage of LEO systems is that their nodes are externally observable: the positions and orbital parameters of satellites are publicly available. We exploit this observability of LEO nodes by conducting a longitudinal study of the Starlink constellation. Our dataset consists of Starlink Two-Line Element (TLE) records spanning 2019 to December 2025 provided by Space-Track~\cite{spacetrack_usspacecom}, which provide snapshots of satellite orbital state, including inclination, right ascension of the ascending node (RAAN), eccentricity, and mean motion. We also use one month of additional TLE data (January 2026) for movement detection. We use these records together with \texttt{sgp4} to derive daily satellite states and reconstruct how the constellation evolves over time. We also incorporate conjunction reports provided by CelesTrak~\cite{celestrak_socrates} to analyze collision avoidance behavior. In addition, we augment the dataset with metadata from Space-Track's Satellite Catalog (SatCat)~\cite{spacetrack_usspacecom}, including satellite launch and decay dates, to support analyses of deployment patterns and satellite lifecycles. When a satellite lacks a TLE for a given day, we use the nearest available record within a short temporal window to maintain consistency in our daily analysis.

%% file: analysis.tex
\section{Analysis}
\label{sec:analysis}

Our analysis aims to extract from Starlink's real deployment the structural properties and modeling parameters most relevant to the networking community. We therefore study Starlink's deployment and operational dynamics through the following research questions:

\begin{itemize}
    \item \textbf{RQ1:} How does the Starlink constellation evolve over time?
    \item \textbf{RQ2:} What is the internal structure of a shell and how does it evolve over time?
    \item \textbf{RQ3:} What are the lifecycle characteristics of satellites, and how frequently do failures occur?
    \item \textbf{RQ4:} What satellite movement patterns occur in practice, and how frequent are they?
\end{itemize}

These questions let us test whether the idealized structures assumed in prior work match the real deployed system and quantify the deviations that are meaningful for LEO network modeling. 

\subsection{Constellation Deployment}

We begin with a macro-level view of the constellation and examine its long-term trends, including the total satellite population, the number of deployed shells, the distribution of satellite generations, and the evolution of satellite's ground visibility. By analyzing the constellation's evolution across multiple dimensions, we address RQ1 in this subsection.

\subsubsection{\textbf{Satellite Population}}

\begin{figure}[h]
\centering
\includegraphics[width=0.7\columnwidth]{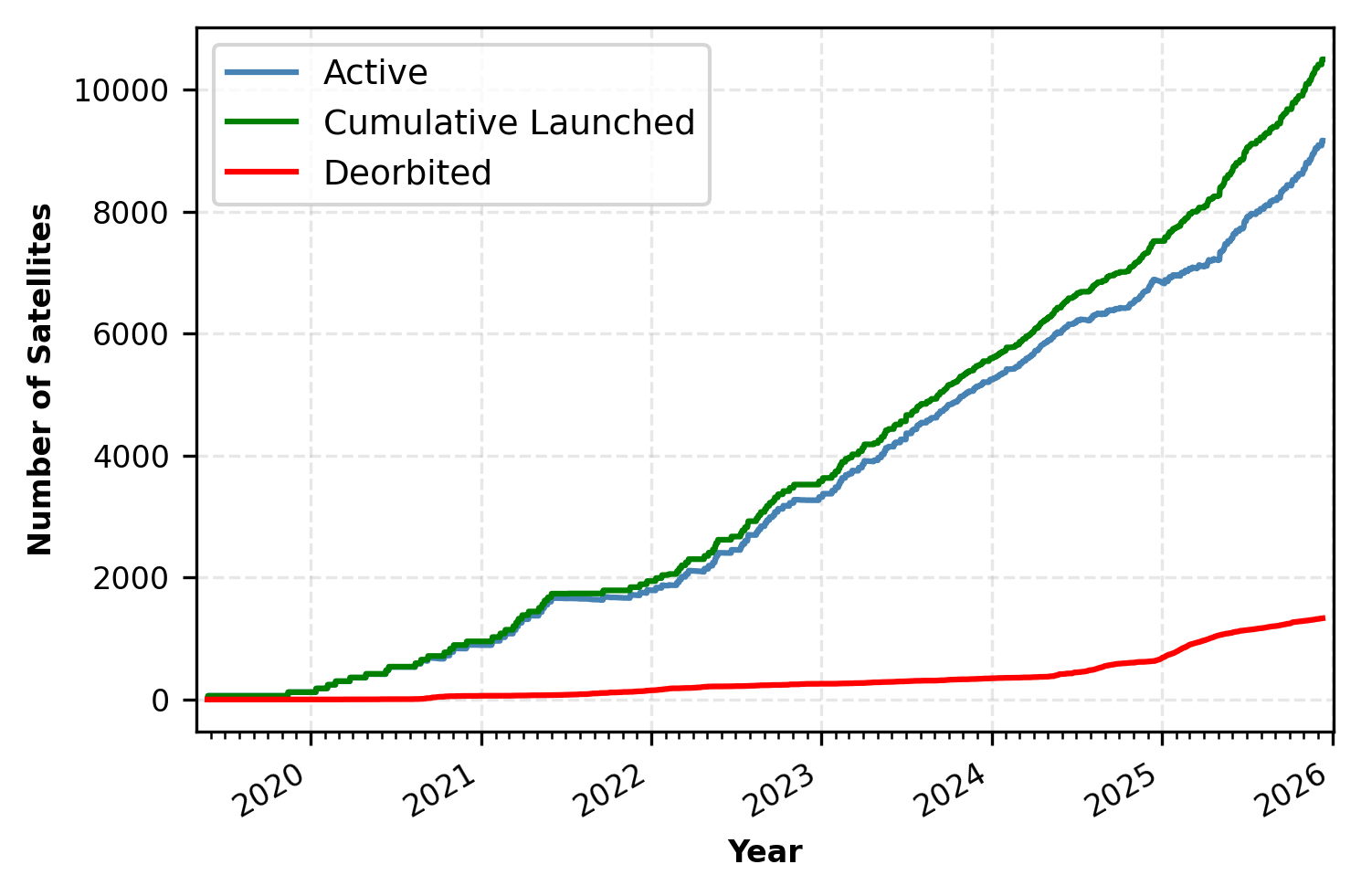}
\caption{Starlink constellation population dynamics}
\Description{Line graph showing three curves from 2019 to 2025: total active satellites reaching 7,500, cumulative deployments reaching 8,500, and cumulative losses exceeding 1,000.}
\label{fig:population_dynamics}
\end{figure}

To contextualize Starlink's deployment scale and pace, we track active satellites in orbit, cumulative launches, and cumulative deorbits over time. Figure~\ref{fig:population_dynamics} shows that deployment began in mid-2019 and reached about 9{,}100 active satellites and roughly 10{,}500 cumulative launches by the end of 2025. The gap between cumulative launches and active satellites reflects deorbits, which exceed 1{,}300 by late 2025. The deorbit curve steepens from 2024 onward, indicating that satellites deployed in 2019--2020 are reaching end of life after roughly five years.

\subsubsection{\textbf{Operational Shells}}

\begin{table}[h]
\centering
\begin{tabular}{ccrrrrr}
\toprule
\textbf{Incl. (°)} & \textbf{Alt. (km)} & \textbf{v1} & \textbf{v1.5} & \textbf{v2 mini} & \textbf{Total} \\
\midrule
53.0 & 552 & 557 & 18  &    0 &  575 \\
53.2 & 540 &   0 & 1376 &    0 & 1376 \\
53.2 & 484 &   0 &   2 & 1846 & 1848 \\
53.2 & 354 &   0 &   2 &  313 &  315 \\
43.0 & 559 &   0 & 625 &  595 & 1220 \\
43.0 & 489 &   0 &   0 & 1380 & 1380 \\
43.0 & 355 &   0 &   0 &  321 &  321 \\
70.0 & 570 &   0 & 352 &  238 &  590 \\
97.6 & 557 &   0 & 208 &  303 &  511 \\
\bottomrule
\end{tabular}
\caption{Identified Starlink orbital shells and satellite generation breakdown as of end of 2025.}

\label{tab:shells}
\end{table}



\begin{figure}[h]
\centering
\includegraphics[width=0.8\columnwidth]{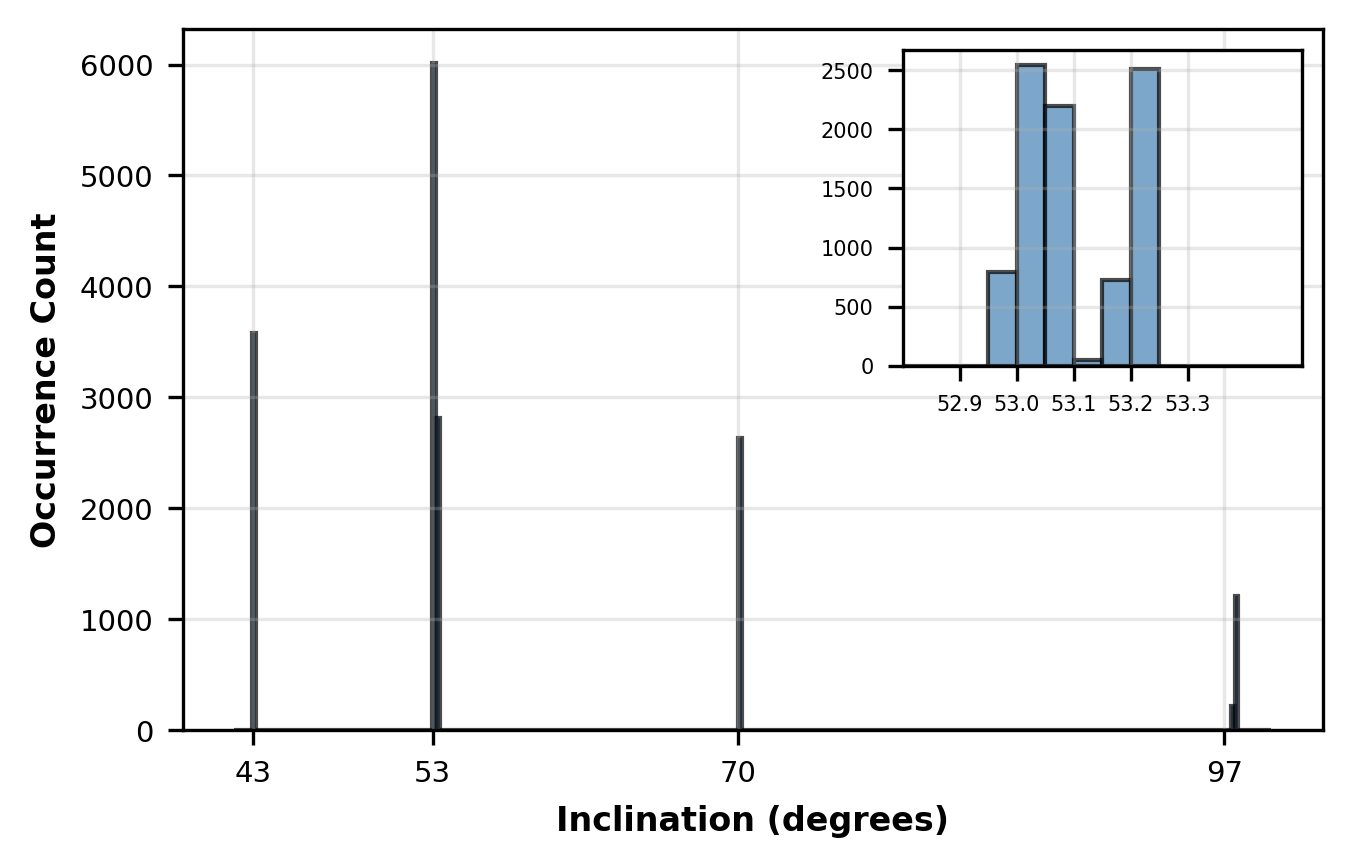}
\caption{Distribution of inclination clusters, with inset highlighting the separation of the 53.0$^\circ$ and 53.2$^\circ$ clusters.}
\label{fig:incl_hist}
\end{figure}

We examine different shells in the constellation (those sharing same inclination and altitude). To identify operational shells directly from TLE data, we apply DBSCAN \cite{10.5555/3001460.3001507} each day with thresholds of $0.1^\circ$ in inclination and 10\,km in altitude and a minimum sample threshold of 25 per shell, determined via elbow analysis over the dataset, to detect persistent shell groupings. This process reveals five distinct inclination bands (Figure~\ref{fig:incl_hist}); within each band, satellites further separate into one or more shells at different altitudes, yielding nine operational shells in orbit currently (Table~\ref{tab:shells}).

\begin{figure}[t]
    \centering
    \includegraphics[width=0.8\columnwidth]{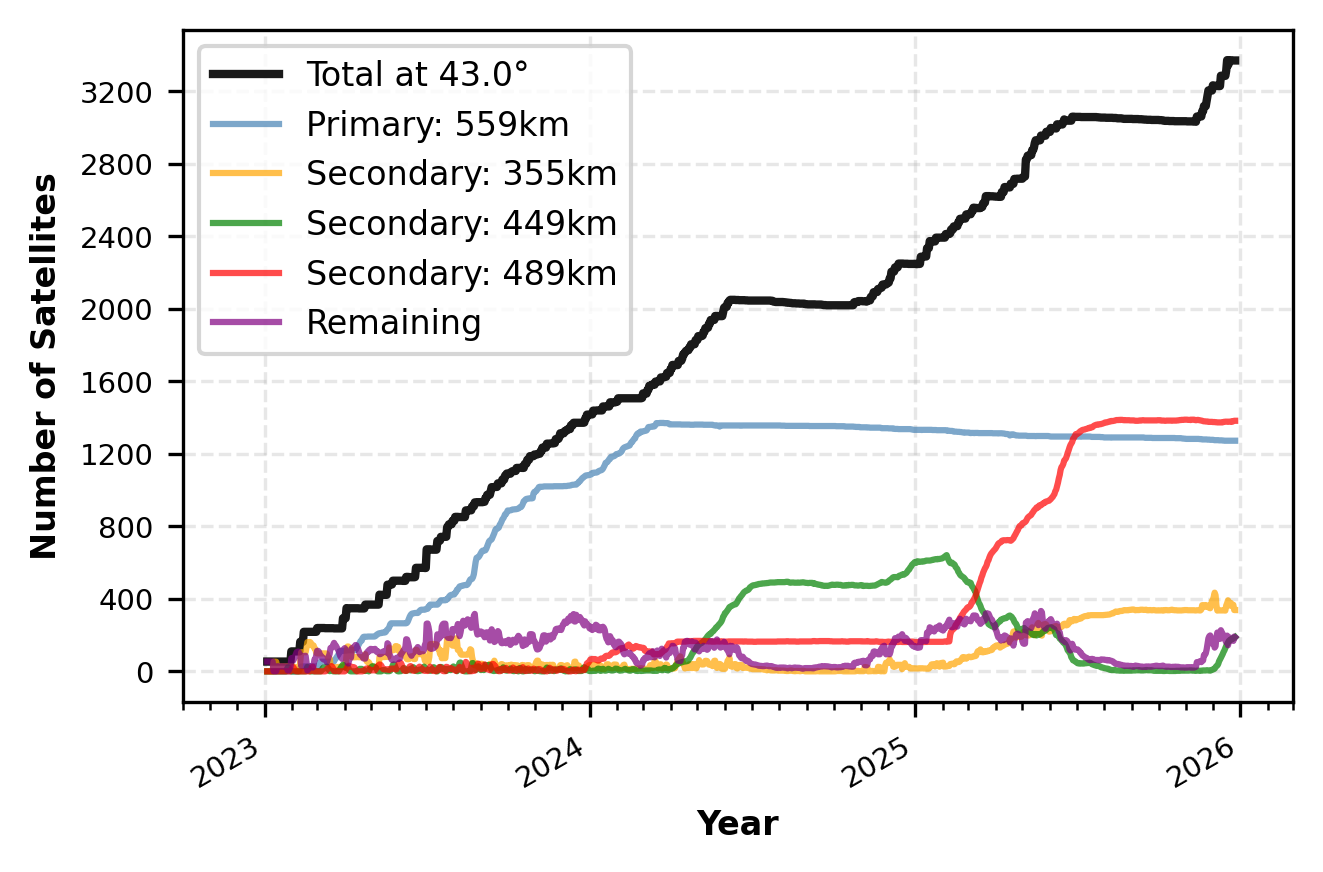}
    \caption{Time-series evolution of satellite populations in the 43.0$^\circ$ inclination group.}
    \label{fig:shell_43}
\end{figure}

\begin{figure}[t]
    \centering
    \includegraphics[width=0.8\columnwidth]{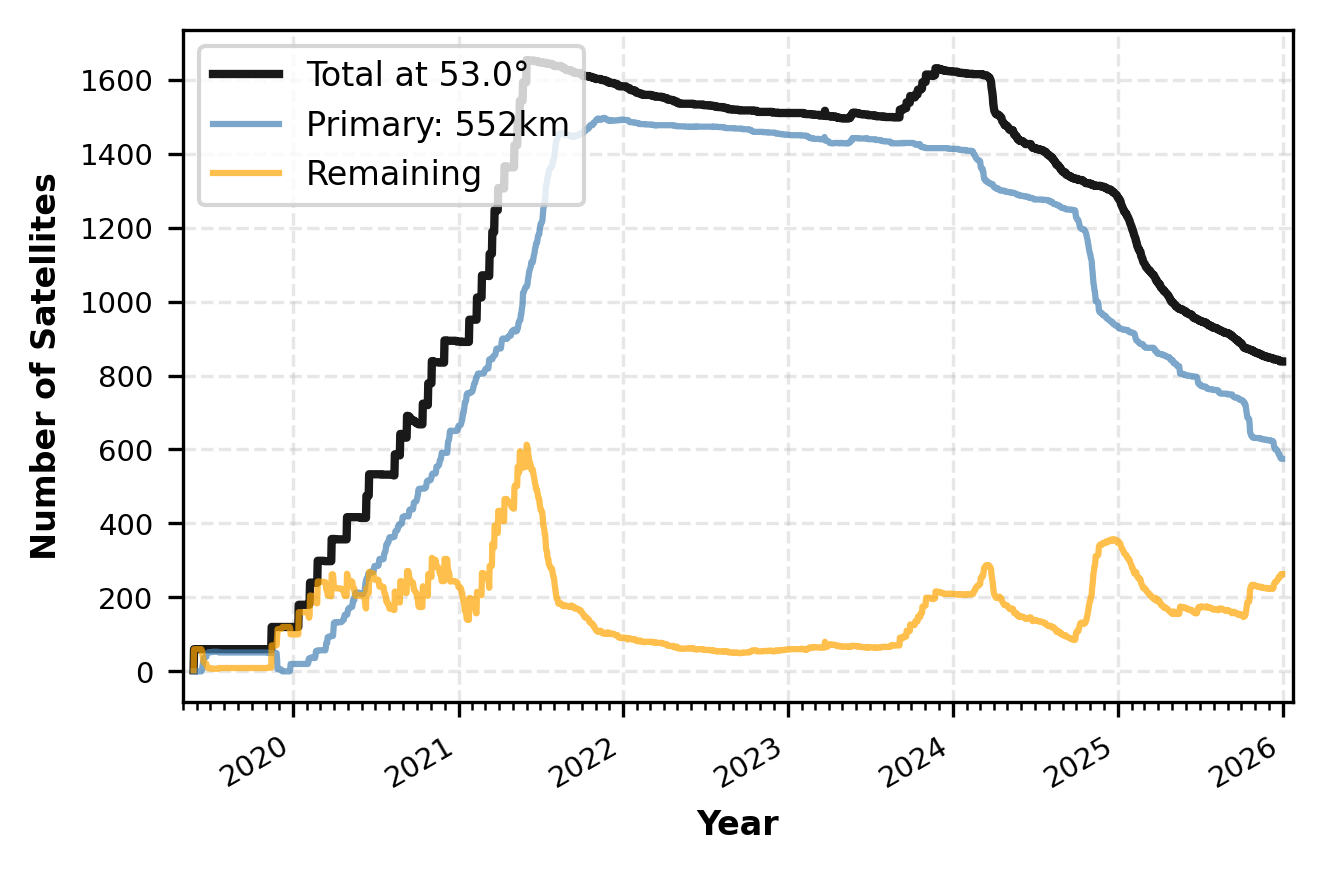}
    \caption{Time-series evolution of satellite populations in the 53.0$^\circ$ inclination group. The remaining satellites are in-transit to the 552km shell.}
    \label{fig:shell_53_0}
\end{figure}

In our analysis, we observe that shell populations evolve over time as launches add satellites and deorbits remove them. Initially, the highest-altitude shell within each inclination group contains the largest population. Figures~\ref{fig:shell_43} and ~\ref{fig:shell_53_0} show this pattern for the 43.0$^\circ$ and 53.0$^\circ$ inclination groups respectively, where the dominant high-altitude clusters (at 559\,km and 552\,km) form the primary operational shells, alongside several lower-altitude secondary shells of varying persistence. Over time, however, this pattern shifts. In the 43.0$^\circ$ inclination group, the lower-altitude 489\,km shell has overtaken the original 559\,km shell as the latter declines due to deorbits. A similar shift appears in the 53.2$^\circ$ inclination group, where the 484\,km shell now contains the most satellites. Interestingly, we also find that some shells are transient. Figure~\ref{fig:shell_43} shows a 449\,km shell in the 43.0$^\circ$ inclination group that appeared after 2024 and disappeared in 2025, lasting only about one year. The same figure also shows an emerging 355\,km shell in the 43.0$^\circ$ inclination group, while Figure~\ref{fig:shell_53_0} shows a similar emerging 354\,km shell in the 53.2$^\circ$ inclination group. Together, these results suggest that Starlink's shell structure remains dynamic, with some shells appearing only briefly and overall deployment shifting toward lower-altitude shells, especially below 500\,km and increasingly below 400\,km.

\begin{figure}[h]
    \centering
    \includegraphics[width=0.8\columnwidth]{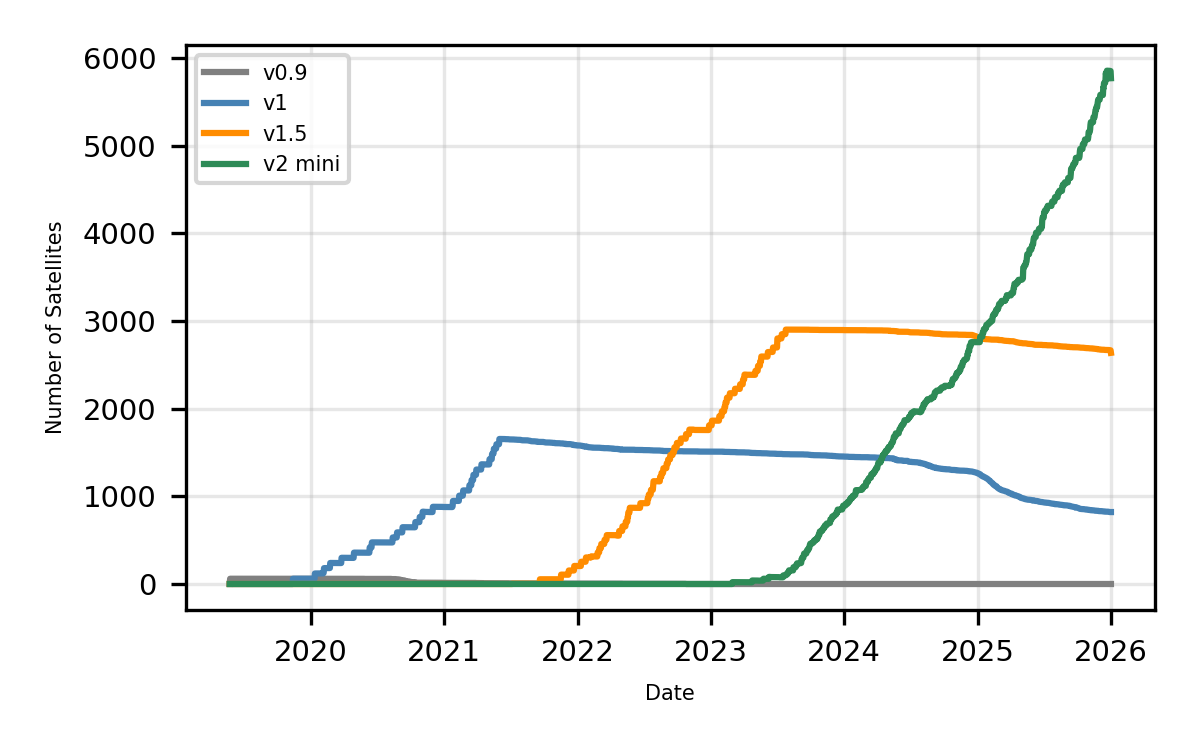}
    \caption{Active satellite counts by generation from 2019 to 2025.}
    \Description{Line plot showing four satellite generations. v1 peaks around 2021 and declines. v1.5 peaks around 2024. v2 Mini grows rapidly from 2023 onward, surpassing all other generations by end of 2025.}
    \label{fig:versions}
\end{figure}

We next examine the different generations of satellites in the constellation. Starlink has deployed four generations of satellites: v0.9 (prototype), v1, v1.5, and v2 Mini. These generations differ in network capabilities and purposes: v0.9 was launched for initial experimentation and deployment and they were deorbited in the same year; v1 satellites lack laser Inter-Satellite Links (ISLs) which were introduced with v1.5 \cite{gunter_starlink_v15}. Figure~\ref{fig:versions} shows the active population of each generation from 2019 to 2025. v1 and v1.5 follow a clear lifecycle --- rapid deployment, a plateau, and eventual decline as newer hardware takes over. v1 peaked at $\sim$1,700 in early 2021 and has been declining since, while v1.5 peaked at $\sim$3,000 in mid-2024 and is now also tapering. v2 Mini, introduced in late 2023, is the dominant and only growing generation by end of 2025, surpassing all predecessors at $\sim$5,800 satellites. As shown in Table~\ref{tab:shells}, newer shells are populated almost exclusively by newer hardware --- the 484 km and 489 km shells are nearly entirely v2 Mini, while the older 552 km and 540 km shells retain v1 and v1.5 satellites respectively. These results show that individual shells contain a mix of satellite generations and therefore exhibit heterogeneous network capabilities.

\subsubsection{\textbf{Comparison with FCC Filings}}

Starlink's deployment plan has evolved across successive FCC filings. After an initial approval near 1{,}000\,km in 2018, the constellation was lowered to about 550\,km and expanded to inclinations of 53.0$^\circ$, 53.2$^\circ$, 70.0$^\circ$, and 97.6$^\circ$ by 2021. Later filings added shells at 525--535\,km for 53$^\circ$, 43$^\circ$, and 33$^\circ$ inclinations~\cite{fcc_gen2_partial_grant}, followed by additional lower-altitude authorizations in 2026, including 480\,km/53$^\circ$ and 485\,km/43$^\circ$~\cite{fcc_low_alt}. Starlink has also proposed lowering the 525--530\,km shells to 484\,km~\cite{foust_2026_lowering}.

We find that the five inclination groups in our data match the initial authorized structure, but the deployed constellation differs from the filings in two ways. First, some shells never reach their filed capacity. For example, the authorized 53.0$^\circ$/552\,km shell has a target size of 1{,}584 satellites, but in practice remains closer to 1{,}500, as shown in Figure~\ref{fig:shell_53_0}. We observe a similar pattern for the 53.2$^\circ$/540\,km shell, which is also planned for 1{,}584 satellites but reaches only about 1{,}500 in the deployed constellation. Second, some observed shells sit at different altitudes than their filed counterparts. For example, we observe a shell at 43.0$^\circ$/559\,km, above the 530\,km altitude authorized in the 2022 partial grant~\cite{fcc_gen2_partial_grant}.

\subsubsection{\textbf{Coverage and Visibility Analysis}}

\begin{figure}[h]
\centering
\includegraphics[width=\linewidth]{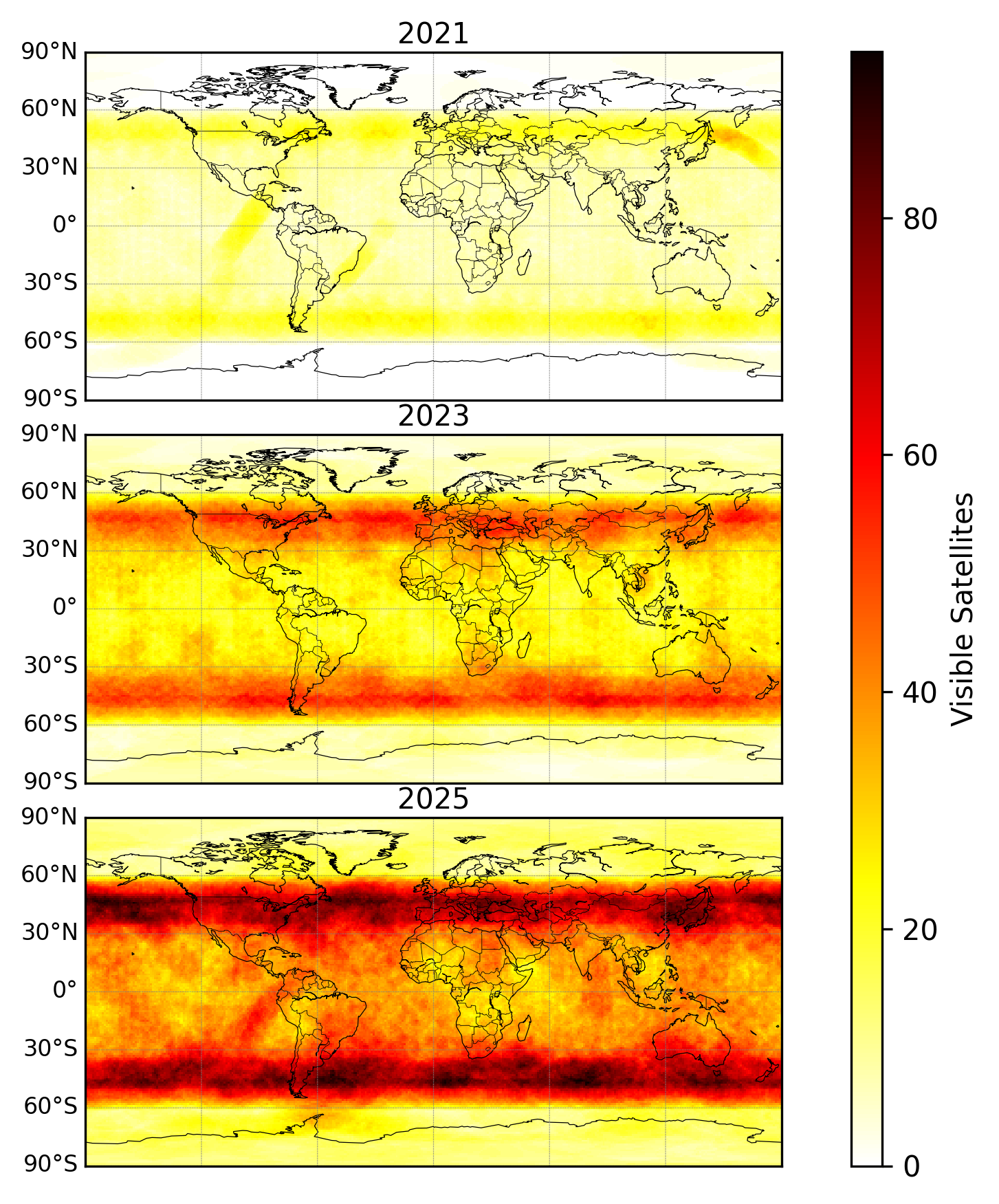}
\caption{Evolution of Starlink satellite visibility for 2021, 2023, and 2025.}
\label{fig:visibility}
\end{figure}

Having characterized the deployed shell structure, we next examine how coverage evolves over time (RQ1). We estimate satellite visibility from ground locations on a 1$^\circ \times$1$^\circ$ grid by counting the number of satellites visible at 12:00 UTC on the last day of each year from 2020 to 2025, using a minimum elevation angle of 25$^\circ$ \cite{fcc2021starlink}. Figure~\ref{fig:visibility} shows representative snapshots for 2021, 2023, and 2025.

Coverage is initially concentrated at mid-latitudes, reflecting early deployment 
of the 53.0$^\circ$/552\,km shell, with most locations having a visibility of about 10--20 satellites. As this shell grows toward peak population, visibility rises to 30--50 satellites across many populated regions by the end of 2023. By 2025, the concurrent expansion of the 43.0$^\circ$, 53.0$^\circ$, and 53.2$^\circ$ shells pushes visibility to roughly 80 satellites across the 30$^\circ$--60$^\circ$N and 30$^\circ$--60$^\circ$S latitude bands, where much of the world's population resides. Notably, coverage is not perfectly symmetric across these dominant latitude bands, and the number of visible satellites still varies across regions.


\subsection{Shell Configuration}

In this subsection, we address RQ2 on shell structure and evolution. We focus on Shell~1 (53.0$^\circ$, 552\,km), the earliest deployed shell and the one with the longest operational history, which makes it a natural reference for analyzing real-world shell dynamics over time. Where useful, we also compare against other shells to highlight broader constellation trends.

\subsubsection{\textbf{Orbital Planes}}

\begin{figure}[t]
    \centering
    \includegraphics[width=0.8\linewidth]{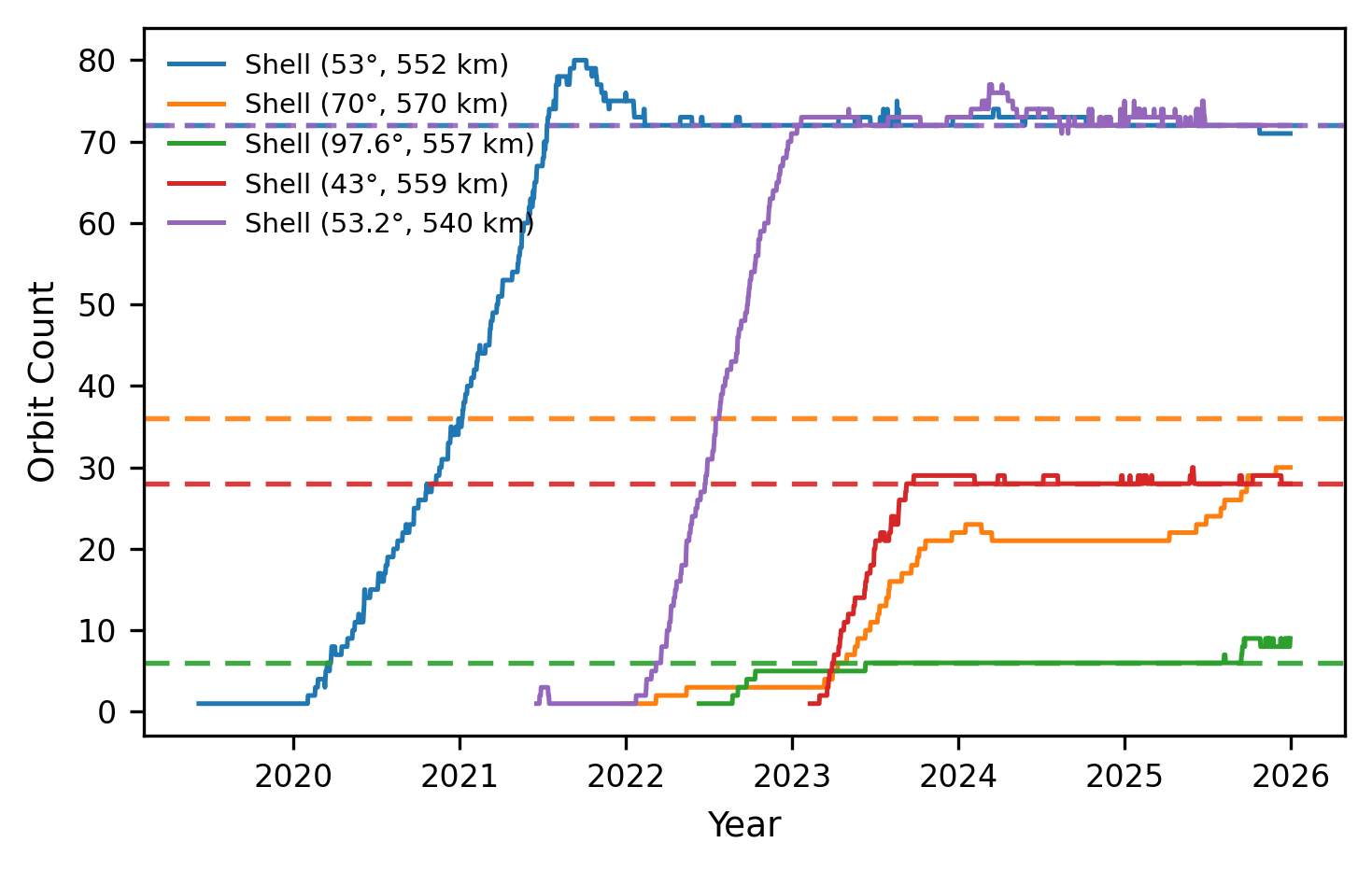}
    \caption{Number of orbital planes observed across the five dominant Starlink shells at different inclinations. Dashed lines denote the nominal plane counts from deployment filings~\cite{fcc2021starlink,fcc_eratum23}.}
    \label{fig:orbit_count}
\end{figure}

To analyze orbital structure, we first cluster satellites into orbital planes using DBSCAN over their RAAN values, since satellites in the same orbital plane have approximately the same RAAN value. Figure~\ref{fig:orbit_count} shows the resulting plane counts over time for the five primary shells, with dashed lines indicating the nominal plane counts reported in FCC filings. As each shell is deployed, the number of detected orbital planes generally increases and then plateaus once deployment matures. However, the  plane count does not always remain below or equal to the ideal value. In several cases, it temporarily exceeds the idealized count. For example, Shell~1, whose filed configuration specifies $72$ planes, reaches as many as $79$ detected planes. Closer inspection shows that these excess planes arise from small intermediate clusters of satellites that form between adjacent nominal planes. Thus, the effective orbital-plane structure of a deployed shell is not strictly fixed by its filing; instead, it evolves over time. 



\subsubsection{\textbf{Intra-orbit Satellite Spacing}}

\begin{figure}[t]
    \centering
    \includegraphics[width=0.8\linewidth]{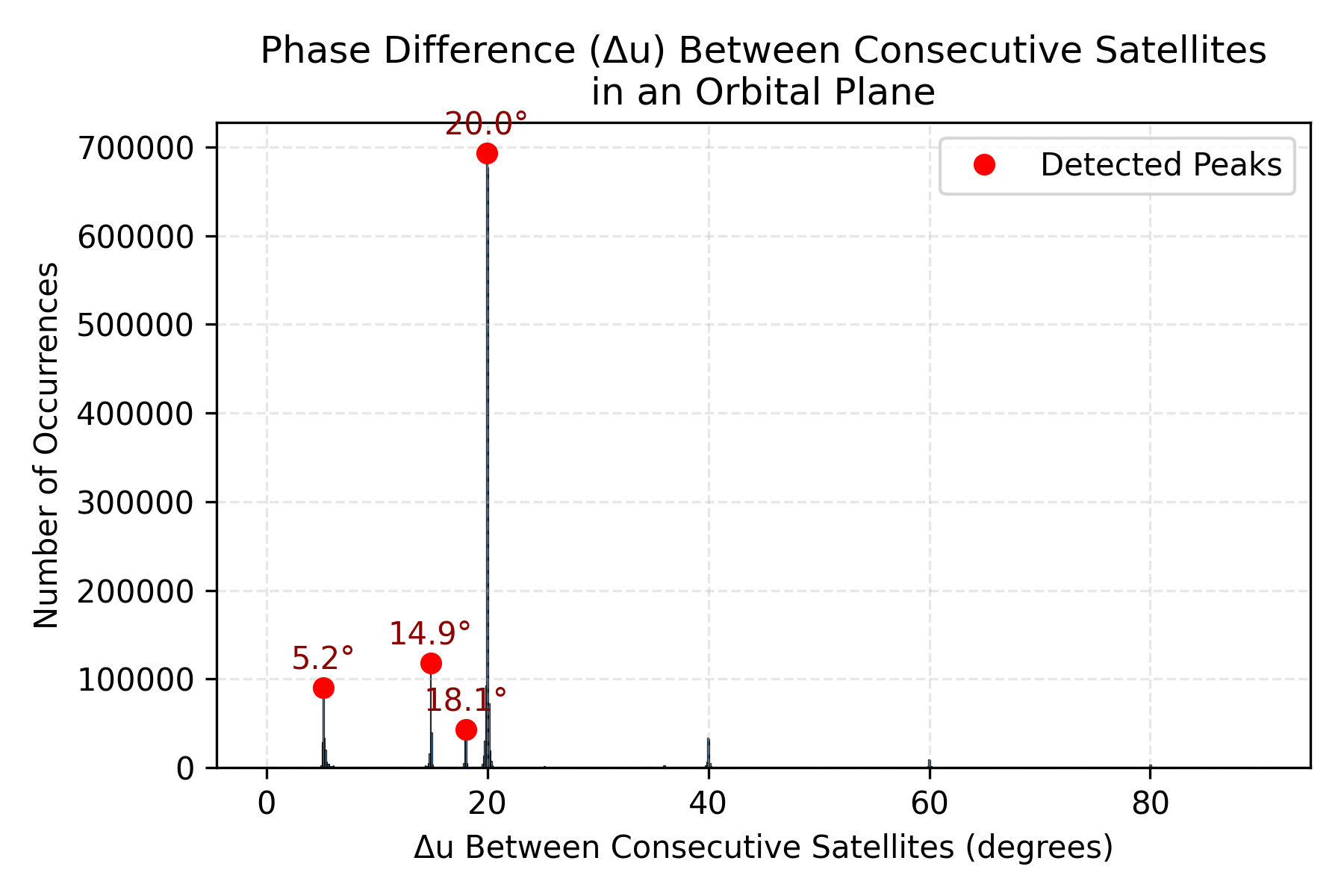}
    \caption{Distribution of phase differences $\Delta(u)$ between consecutive satellites within an orbit in Shell~1.}
    \label{fig:ta_diffs_hist}
\end{figure}

In an ideal Walker-Delta constellation, satellites within an orbital plane are evenly spaced in phase. To examine whether this property holds in real deployments, we analyze the phase spacing $u$ between consecutive satellites within each orbit. Figure~\ref{fig:ta_diffs_hist} shows the distribution of phase differences $\Delta(u)$ aggregated across the entire observation period in Shell-1. The distribution exhibits two clear peaks near $20^\circ$ and $5^\circ$. The dominant peak around $20^\circ$ defines a set of satellites that are spaced nearly uniformly along the orbit. We call them \emph{regular satellites}. This corresponds to about  18 regular satellites per orbital plane ($360^\circ / 18 = 20^\circ$) and forms the regular structure of the plane. In contrast, the smaller peak near $5^\circ$ reflects satellites that are much more closely spaced.

\begin{figure}[h]
    \centering
    \includegraphics[width=0.7\linewidth]{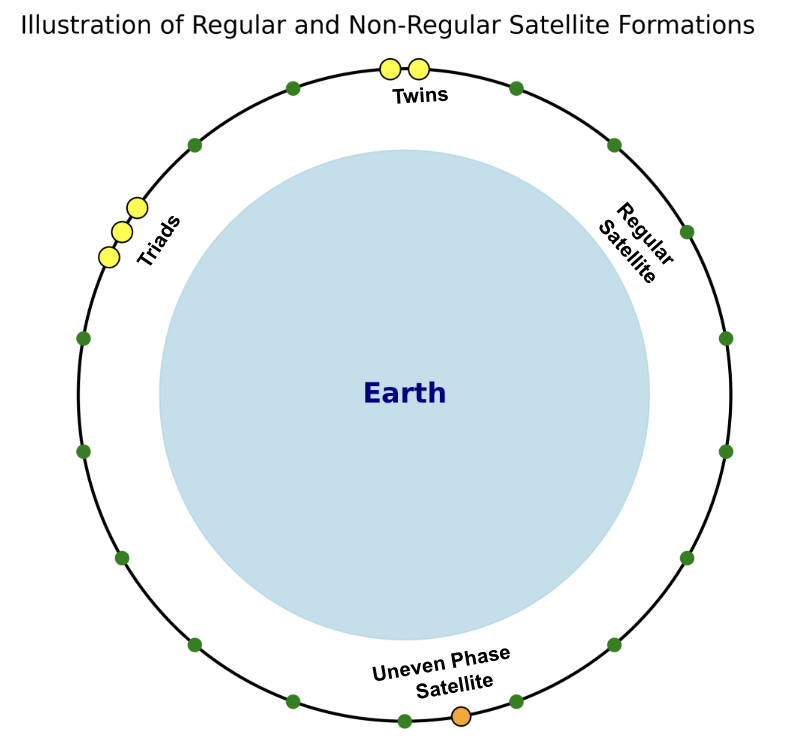}
\caption{Illustration of regular satellites, uneven non-regular satellites, and clustered formations  \emph{twins} and \emph{triads} within the same orbit.}    \label{fig:regular_non_regular}
\end{figure}

We classify satellites that do not follow the regular spacing as \emph{non-regular satellites}. These fall into two groups. The first consists of tight clusters that produce the peak near $5^\circ$ in Figure~\ref{fig:ta_diffs_hist}; we call these \emph{twins} and \emph{triads} when two or three satellites lie within $5^\circ$ of each other. The second consists of satellites at uneven phase positions that do not belong to such clusters. Figure~\ref{fig:regular_non_regular} illustrates this distinction: regular satellites (green) follow the periodic phase distance, while non-regular satellites appear either as twin/triad clusters (yellow) or as unevenly placed satellites (orange). We hypothesize that non-regular nodes (twins, triads, and uneven phase satellites) are acting as backup nodes of the shell.

We next compare the observed regular spacing with the ideal spacing implied by the FCC filings. We find out that distinction between \emph{regular} and \emph{non-regular} satellites appears consistently across the other primary Starlink shells as well. Table~\ref{tab:regular_spacing} summarizes the regular spacing measured in five major shells and shows that, in every case, the observed spacing differs from the ideal value derived from the original FCC design. This indicates that deviations from the filed configuration are a general feature of the real deployment rather than an isolated property of Shell~1. We further observe close satellite clusters  (twin and triad satellites) in all shells. Their characteristic spacing is summarized in Table~\ref{tab:twin_spacing}.

\begin{table}[t]
\centering
\small
\resizebox{\linewidth}{!}{%
\begin{tabular}{lcccc}
\hline
\textbf{Shell} &
\textbf{Regular} &
\textbf{Authorized} &
\textbf{Regular} &
\textbf{Authorized} \\
\scriptsize (inclination, altitude) &
Spacing &
Spacing &
Sats/Plane &
Sats/Plane \\
\hline
53, 552   & 20.0 & 16.4 & 18 & 22 \\
70, 570   & 22.6 & 18.0 & 16 & 20 \\
97.6, 557 & 10.0 & 8.4 & 36 &  58 \\
43, 559   & 6.0  & 3.0 & 60 & 120 \\
53.2, 540 & 20.0 & 16.4 & 18 & 22 \\
\hline
\end{tabular}}
\caption{Observed regular phase spacing of five primary shells compared with authorized spacing derived from constellation filings~\cite{fcc2021starlink,fcc_eratum23}.}
\label{tab:regular_spacing}
\end{table}

\begin{table}[h]
\centering
\begin{tabular}{lc}
\hline
\textbf{Shell \scriptsize (inclination, altitude)} & \textbf{Twin/Triad Spacing \scriptsize (deg)} \\
\hline
53, 552   & 5.2 \\
70, 570   & 0.9 \\
97.6, 557 & 1.4 \\
43, 559   & 1.6 \\
53.2, 540 & 5.0 \\
\hline
\end{tabular}
\caption{Observed phase separation of clustered satellites forming twins and triads in five primary shells.}
\label{tab:twin_spacing}
\end{table}

\begin{figure}[h]
    \centering
    \includegraphics[width=0.95\linewidth]{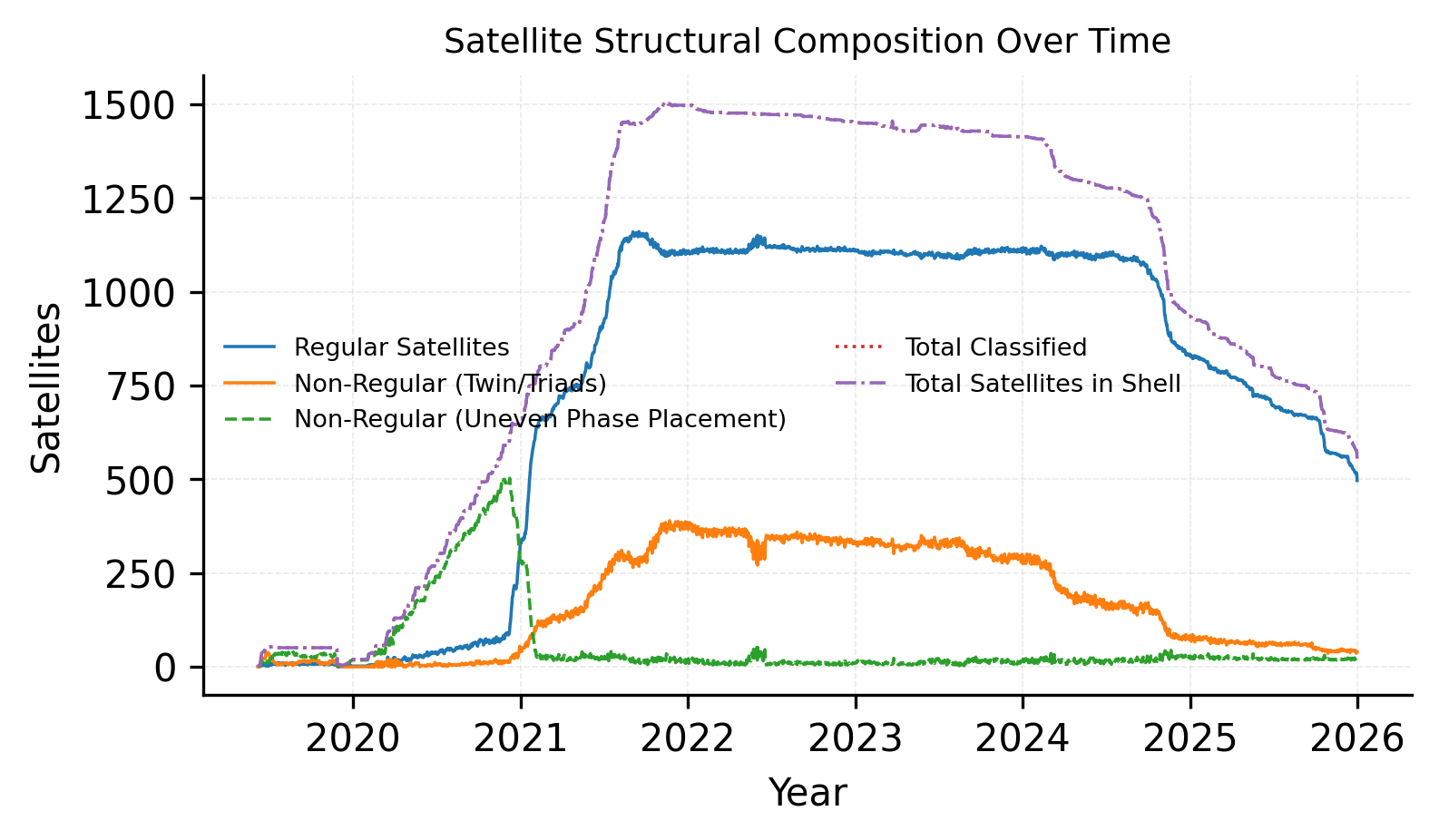}
    \caption{Structural composition of Shell-1 satellites over time, showing the breakdown into regular and non-regular satellites.}
    \label{fig:reg_shell_1}
\end{figure}

To understand how regular and non-regular satellites evolve over time, we examine the structural composition of Shell~1. Figure~\ref{fig:reg_shell_1} shows its evolution over the observation period. Early in deployment, the shell consists of a mix of regular and non-regular satellites as the constellation undergoes initial deployment and reconfiguration. By mid-2020 to 2021, the shell settles into a more stable structure, with regular satellites emerging as the dominant group. After this point, approximately 75--80\% of satellites belong to the regular structure. Clustered formations such as twins and triads form the second-largest group, while only a small fraction of satellites remain irregularly phased relative to the regular backbone.

\subsubsection{\textbf{Regular satellite stability.}}

We next examine whether the regular phase spacing remains stable over time. We find that the spacing $\Delta(u)$ between regular satellites in Shell~1 changes substantially during deployment. It is about $6^\circ$ in 2019 during the initial deployment phase, then increases to roughly $18^\circ$ in 2020 and remains at that level for about a year, and then changes again and finally stabilizes to $20^\circ$ from 2021 onward. These results show that the shell's internal regular structure does not remain fixed, but instead evolves significantly over time.

\begin{figure}[t]
    \centering
    \includegraphics[width=0.7\columnwidth]{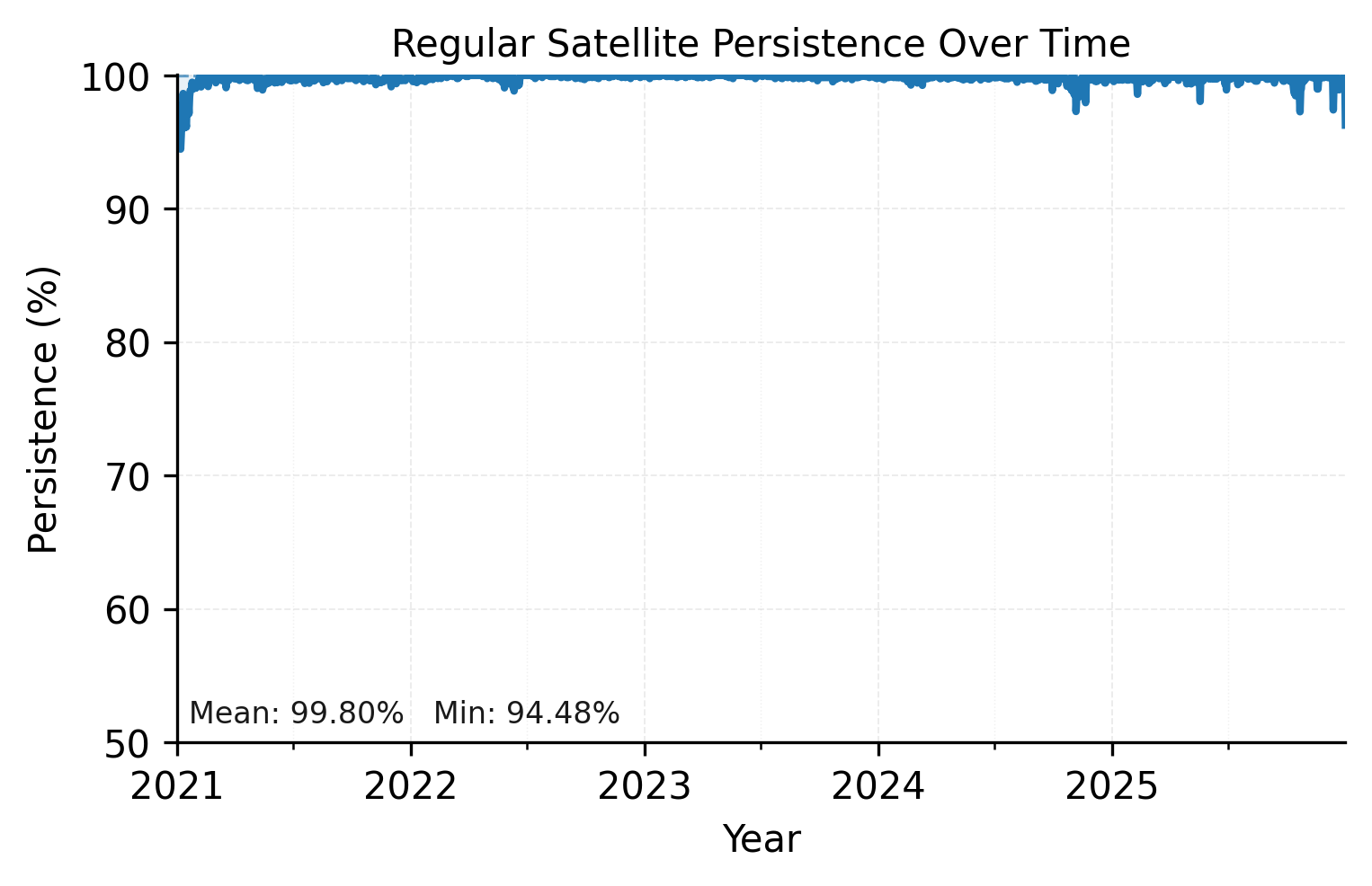}
    \caption{Daily regular persistence across entire shell in Starlink Shell--1. Regular satellites are highly persistent across consecutive days.}
    \label{fig:regular_stability}
\end{figure}

Because the regular spacing stabilizes in 2021, we focus on the period from 2021 onward to examine whether the satellites forming this structure also remain stable over time. We first measure day-to-day persistence by tracking what fraction of satellites classified as regular on one day remain regular on the next. As shown in Figure~\ref{fig:regular_stability}, this fraction stays close to $100\%$ throughout the observation period, with only small and short-lived drops. This indicates that once a satellite becomes part of the regular structure, it is very likely to remain regular on the following day.

\begin{figure}[t]
    \centering
    \includegraphics[width=0.7\columnwidth]{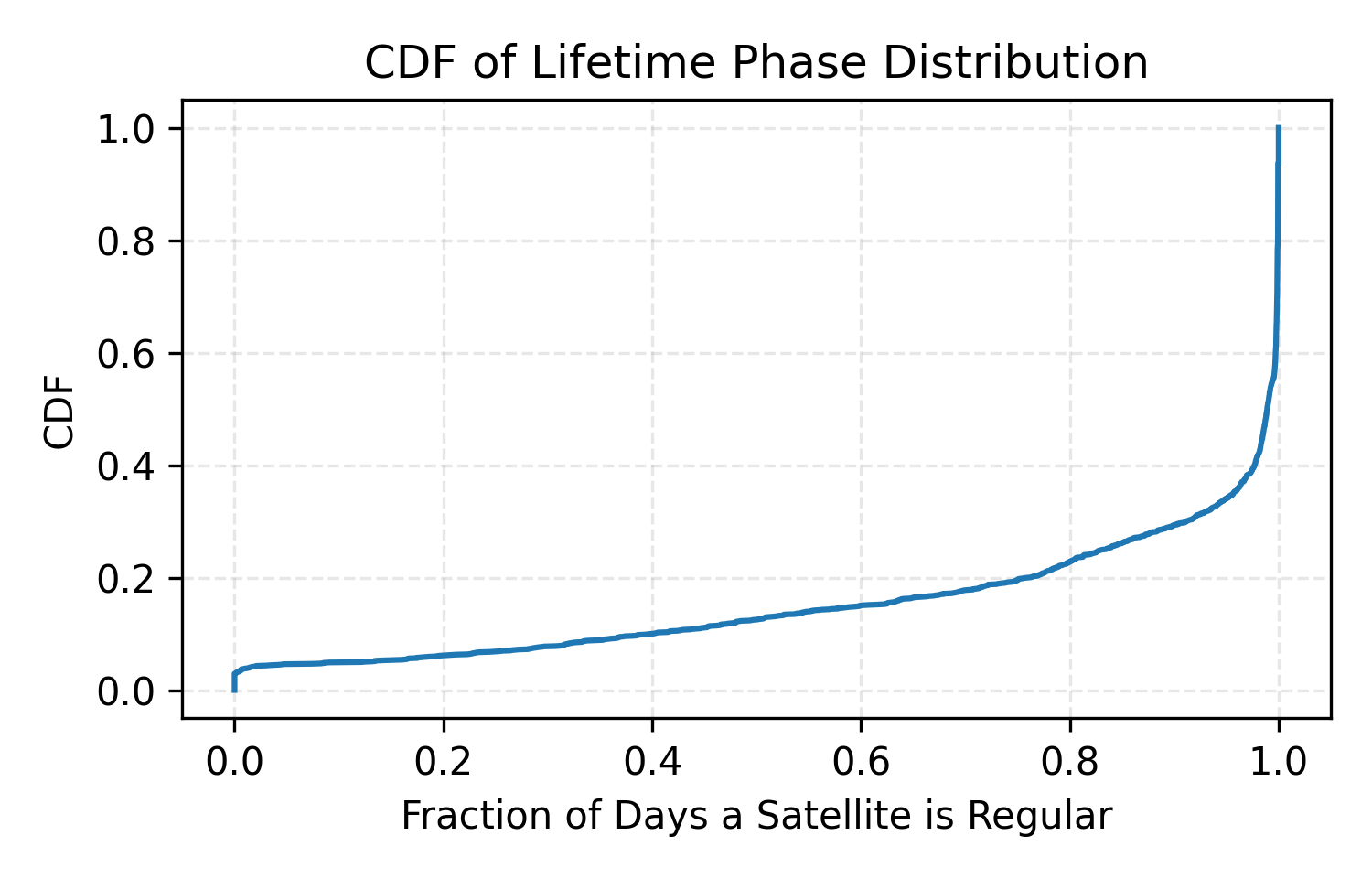}
    \caption{Per-satellite regular lifetime fraction in Starlink Shell--1. Many satellites are classified as regular for only part of their observed lifetime.}
    \label{fig:regular_fraction}
\end{figure}

At the same time, satellites do not necessarily remain regular throughout their full observed lifetime. For each satellite, we measure the fraction of observed days on which it is classified as regular. Figure~\ref{fig:regular_fraction} shows that many satellites are regular for only part of their observed lifetime, with only about $60\%$ remaining regular throughout all observed days. This indicates that satellites can shift between regular and non-regular phase positions over time.

Overall, our results show that: (a) the regular phase spacing evolves over time, (b) once the regular spacing stabilizes, the regular structure remains highly stable from day to day, and (c) individual satellites can still transition between regular and non-regular positions over longer periods.


\subsubsection{\textbf{Walker-Delta Parameters}}

A Walker-Delta constellation assumes satellites are evenly spaced within each orbital plane. In Shell~1, this property is observed in the subset of \emph{regular satellites}. We therefore map the deployed Shell~1 structure to Walker-Delta parameters by focusing only on these regularly placed satellites. Based on the measured regular spacing of approximately $20^\circ$, Shell~1 contains $18$ regular satellites per plane across $72$ orbital planes, giving a total of $T=1296$ regular satellites.

To estimate the Walker phasing parameter $F$, we analyze representative snapshots from 2022, 2023, and 2024, when Shell~1 exhibits a mature and largely regular structure. For each regular satellite in orbital plane $p$, we identify the regular satellite in the adjacent plane $p+1$ with the smallest difference in phase $u$. This gives an estimate of the inter-plane phase offset $\Delta\phi$. Using the Walker-Delta relation
\[
\Delta\phi = \frac{360F}{T},
\]
we compute the corresponding $F$ values across adjacent plane pairs. Across all evaluated snapshots, the dominant estimate is $F=45$, which corresponds to an inter-plane phase shift of approximately $\Delta\phi \approx 12.5^\circ$. The effective Walker-Delta parameters of the regular structure in Shell~1 can therefore be written as
\[
\text{Walker-Delta: } i: T/P/F = 53:1296/72/45.
\]

This differs from the originally planned Shell~1 configuration of $53:1584/72/F$, where $F$ is the phasing parameter chosen by the operator and not explicitly reported in FCC filings. 


\subsection{Satellite Life Cycle}

To address RQ3, we next analyze satellite lifecycles. Because LEO constellations evolve continuously under both operational decisions and real-world effects, satellite lifetimes vary substantially: some deorbit within months of deployment, while others remain in orbit for more than six years, exceeding the nominal five-year design life. To understand this variation, we examine both overall satellite lifespan and the durations of different lifecycle phases.

\begin{figure}[h]
    \centering
    \includegraphics[width=0.7\linewidth]{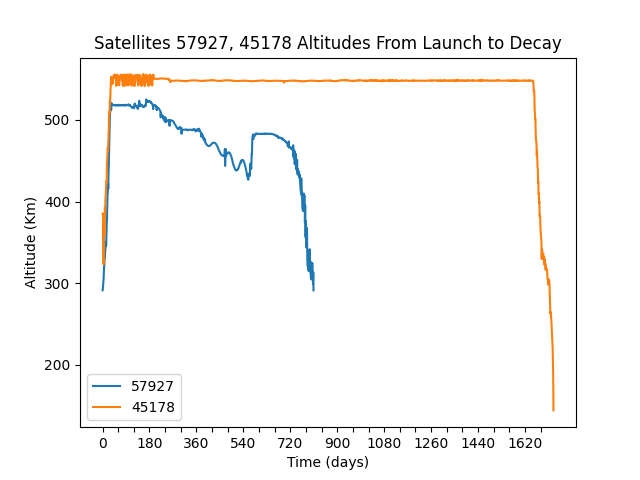}
    \caption{Example of satellite lifetimes.}
    \label{Example lifetimes}
\end{figure}

\subsubsection{\textbf{Lifecycle Phases}}

We classify each satellite's lifecycle into three distinct phases. The ascent 
phase begins at the satellite's first TLE appearance and ends when it first 
reaches within 10 km of its operational altitude. The operational phase spans 
from this first entry into the operational shell to the last exit. The descent 
phase covers the time from last shell exit to the satellite's final TLE 
observation. The 10 km margin was chosen to distinguish true phase transitions 
from routine station-keeping maneuvers while remaining tight enough to capture 
actual altitude changes.

Figure~\ref{Example lifetimes} illustrates these phases for two representative 
satellites. Satellite 45178 (orange) follows the typical pattern: a brief ascent 
to operational altitude, a stable operational phase lasting over 1,600 days, 
followed by a controlled descent. In contrast, satellite 57927 (blue) deorbits 
after roughly 850 days, with notable altitude excursions during its operational 
phase suggesting active maneuvering prior to early deorbit.

Figure~\ref{fig:lifecycle} (in the appendix) shows the population-level duration 
of each phase across satellites in our dataset that have reached their operational altitude. The ascent phase has a 
median of 64 days (mean 77.9 days), with most satellites reaching operational 
altitude within 40--80 days. The tight distribution reflects consistent deployment 
procedures, though some satellites take over 300 days due to delayed orbit raising. 
The operational phase varies widely, with a median of 864 days (mean 901.4 days), 
likely reflecting different satellite batches or generations with different planned 
lifetimes. The descent phase is more concentrated, with a median of 107 days 
(mean 148.0 days), with most satellites completing deorbit within 50--150 days 
of leaving operational altitude.

\subsubsection{\textbf{Lifetime distribution}}
\begin{figure}[h]
    \centering
    \includegraphics[width=0.7\linewidth]{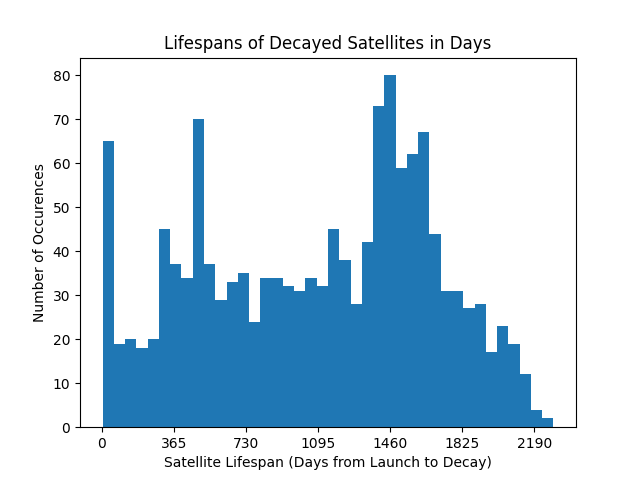}
    \caption{Lifetime distribution of decayed satellites}
    \label{Lifespan of decayed satellites}
\end{figure}

To analyze satellite reliability, we examine the lifetime distribution of all 
decayed satellites. Figure~\ref{Lifespan of decayed satellites} shows that the 
distribution roughly resembles a bathtub curve: an elevated early failure rate, a 
relatively flat  operational phase  region, and a rising wear-out peak for satellites reaching expected end-of-life. To further investigate this, we apply Kaplan-Meier (KM) survival analysis~\cite{kaplan1958nonparametric}. For each satellite, we compute its operational lifetime as the duration between its first and last appearance in the TLE dataset. The KM estimator computes the survival function $S(t)$, the probability that a satellite remains operational beyond day $t$, by iteratively updating survival probability at each observed deorbit event:
where $d_i$ is the number of deorbits and $n_i$ is the number of satellites at risk at time $t_i$, computed as $S(t) = \prod_{t_i \leq t} \left(1 - \frac{d_i}{n_i}\right)$. 
From $S(t)$, we get the failure probability $F(t)$, the probability that a satellite will fail on a given day t, conditioned on survival upto $t$: $F(t) = \frac{-\Delta S(t)}{S(t) \cdot \Delta t}$
\begin{figure}[h]
    \centering
    \includegraphics[width=\linewidth]{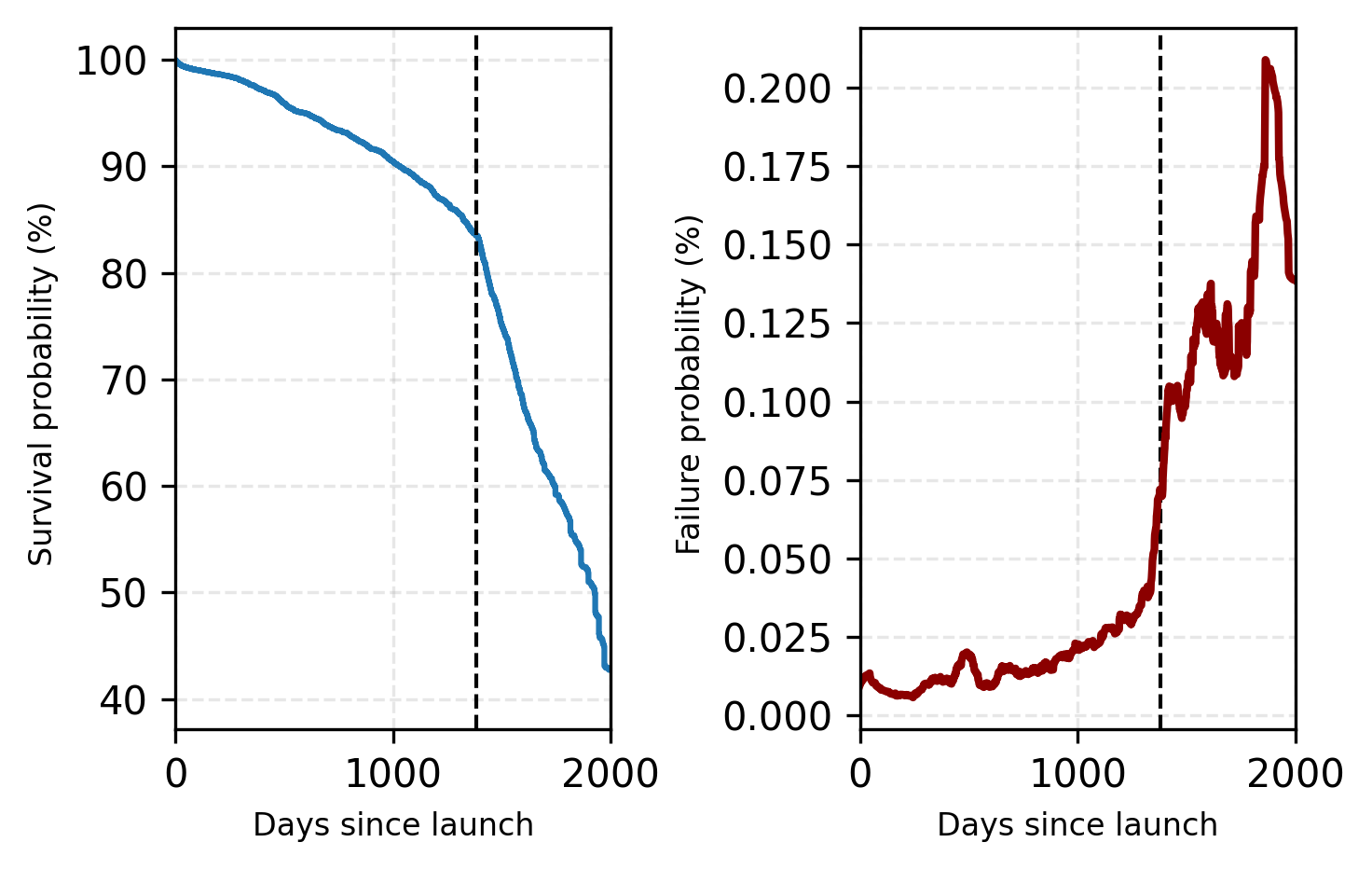}
    \caption{Survival probability and failure probability}
    \label{fig:km}
\end{figure}
Figure~\ref{fig:km} shows the resulting survival and failure probability curves. 
Satellites maintain high reliability through the early operational period, with 
$S(t) \approx 0.83$ at day 1,395. The failure probability $F(t)$ peaks around day 1,850, 
indicating a concentrated wear-out phase as satellites approach end-of-life. 
The failure probability then declines as the remaining population consists of 
newer satellites yet to reach this age. A significant fraction of satellites 
in our dataset remain operational, so survival estimates beyond day 1,395 are 
based on a smaller at-risk population. To characterize the baseline failure 
rate during normal operations, we focus on the first 1,000 days --- prior to 
any visible onset of wear-out in the failure curve --- yielding a mean daily 
failure probability of 0.0128\%. This baseline failure probability reflects true in-orbit failures --- satellites that deorbit unexpectedly --- rather than planned 
end-of-life deorbits, which begin to dominate only after day 1,395.

Together, these results address RQ3: satellite failures are rare during normal 
operations but accelerate sharply in a concentrated end-of-life window, with most 
satellites operating reliably for approximately 4--6 years.

\subsection{Satellite Movement}
To address RQ4, we study satellite movement patterns within operational shells. Satellites within the same operational shell share the same inclination and altitude, and thus experience similar natural perturbations and nearly uniform day-to-day drift. Prior work detects Starlink movements by propagating each satellite's trajectory with SGP4 and comparing predicted and observed orbital parameters~\cite{CollisionAvoidanceStarlink}. This is effective in identifying whether a satellite's trajectory changes, but it has two key limitations. First, it uses each satellite's own propagated path as the reference, making it difficult to determine whether an observed change is significant relative to the rest of the shell. As a result, maneuvers that are already embedded in the latest TLEs may be absorbed into the propagated trajectory and appear nominal on the following day. Second, it focuses primarily on altitude changes, and can therefore miss in-orbit phase adjustments that reposition satellites without a large altitude excursion.

We address these limitations by designing a shell-relative movement detection method. For each shell and each orbital parameter of interest, such as altitude, phase, or RAAN, we first compute the day-to-day change for all satellites in the shell and then estimate the shell's typical change using the median. We then measure how much each satellite deviates from this shell-wide median, allowing us to identify abnormal behavior relative to the population rather than relative only to its own projected path. This yields a more complete empirical view of Starlink's movement patterns and captures both altitude maneuvers and in-plane phasing changes. 
\subsubsection{\textbf{Movement Detection}}





For each shell, we compare satellite states across two consecutive days using a common phase-aligned reference. We set the first reference time to 12:00 UTC on day 1, then define the next reference time on each subsequent day by advancing in integer multiples of the shell's mean orbital period until reaching a time close to 24 hours later. Across days, this aligns satellites by orbital phase ($u$) rather than by fixed clock time. This gives us consistent comparison points across days.

For each satellite $i$, we then compute its day-to-day changes in phase, apogee altitude, and RAAN, denoted by $\Delta u_i$, $\Delta h_i$, and $\Delta \Omega_i$, respectively. For each parameter, we compute the shell-wide median $m$ and standard deviation $s$, where $s$ measures the typical spread of day-to-day changes across satellites over the two days. A satellite is flagged as moving if its change is unusually large relative to this spread, that is, if $|x_i - m| \geq N_\sigma s$ for any parameter $x_i \in \{\Delta u_i, \Delta h_i, \Delta \Omega_i\}$. We use $N_\sigma=5$ as a conservative threshold to avoid misclassifying ordinary day-to-day variation as true movement. Since small changes in phase, altitude, and RAAN can arise from nominal orbital drift, estimation noise, or minor fluctuations shared by many satellites, a lower threshold would flag too many satellites that still follow the shell's normal behavior. 
%

\begin{figure}[h]
    \centering
    \includegraphics[width=0.8\columnwidth]{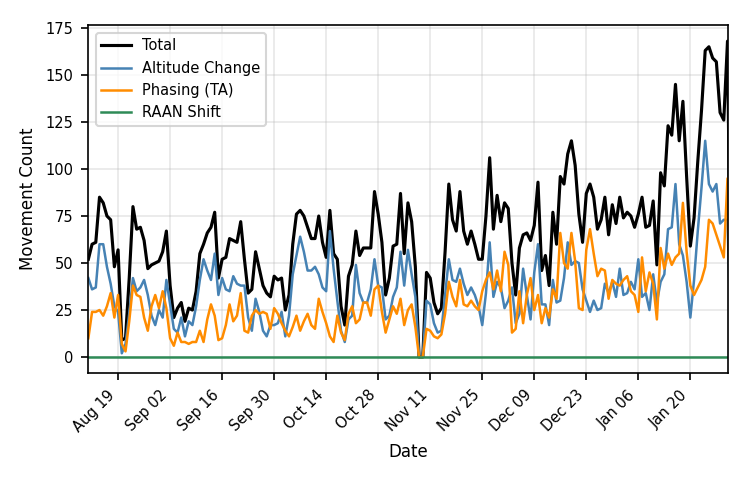}
    \caption{Detected movements per day for the 53.2°, 484 km shell}
    \Description{Time series plot showing daily movement counts from August 2025 to January 2026, broken down into altitude change and phasing categories.}
    \label{fig:maneuvers_53_484}
\end{figure}

We first apply our detection methodology to the 53.2$^\circ$/ 484 \, km shell, which 
contains the largest satellite population in our dataset (up to 2,140 satellites 
on a given day). Figure~\ref{fig:maneuvers_53_484} shows detected movements per 
day over the 174-day study period (August 2025 through January 2026), broken down 
by type. Altitude changes dominate throughout, with phasing movements contributing 
a secondary but sustained component. No RAAN-based ($\Omega$) movements are detected, 
indicating an absence of significant orbital plane-change movements. Movement activity increases 
steadily over the study period, consistent with the ongoing growth of this shell. Scaling to five major Starlink shells (43.0° -- 489 km, 559 km; 53.0° -- 552 km; 53.2° -- 484 km, 540 km), covering an average of 5,948 satellites per day, we detect 45,945 movements over the study period, averaging $\sim$264 movements per day.

We also examined whether phasing maneuvers are consistently preceded by a detectable altitude change. We find that $\sim$92\% of detected phasing movements show no clearly observable accompanying altitude change, suggesting that the altitude component of such maneuvers often falls below our detection threshold, or that some phasing adjustments are made independently of any altitude change altogether. This justifies treating phasing deviation as a separate movement detection parameter.

\subsubsection{\textbf{Movement Duration and Frequency}}

\begin{figure}[h]
    \centering
    \includegraphics[width=0.7\columnwidth]{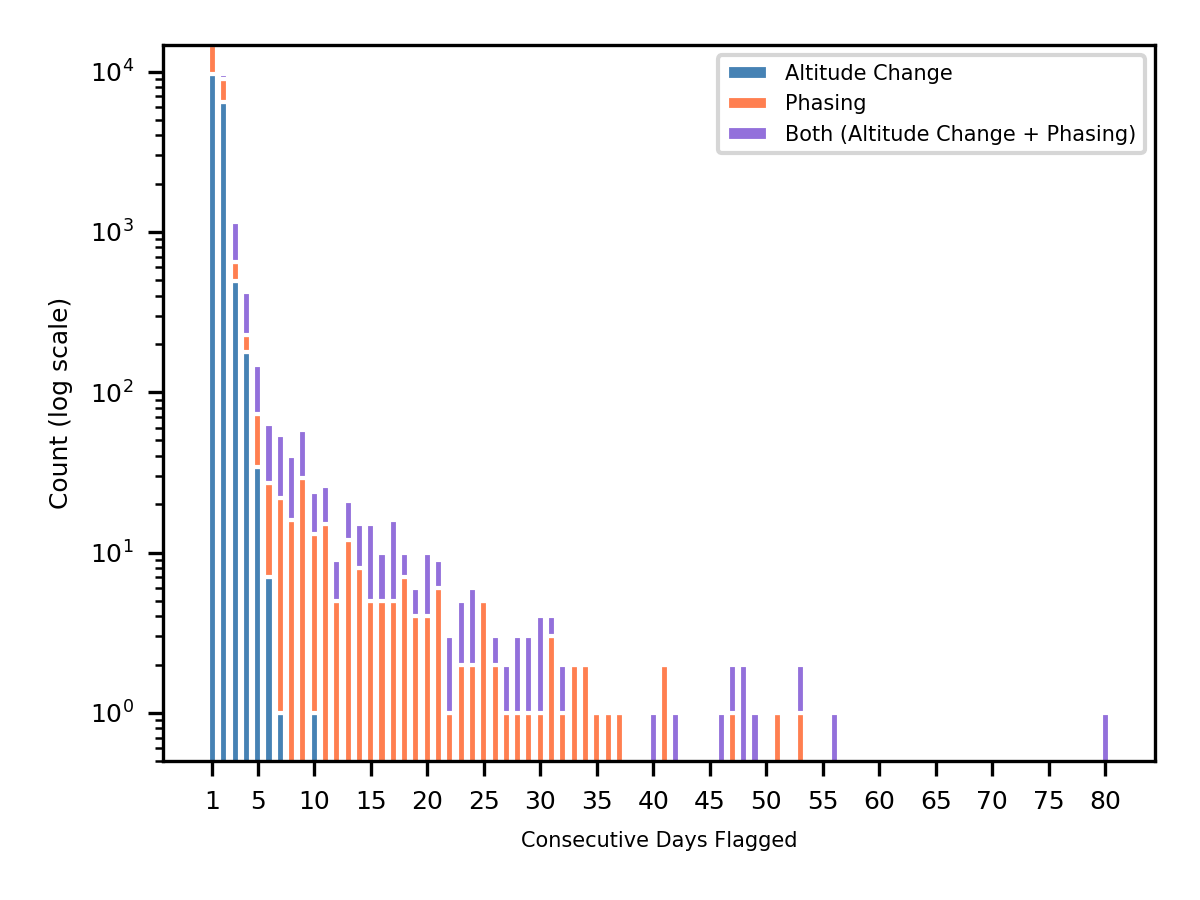}
    \caption{Movement length distribution across major shells.}
    \Description{Bar chart showing the distribution of consecutive days flagged per movement streak, broken down by type. Single-day streaks dominate, with a long tail extending to 80 days.}
    \label{fig:movement_length_all}
\end{figure}

To characterize movement behavior, we group consecutive movements per satellite and analyze their length distribution. Figure~\ref{fig:movement_length_all} shows the distribution across major shells on a log scale. Altitude-change movements concentrate in the 1--5 day range, consistent with short targeted adjustments, while phasing and combination movements dominate the longer tail, reflecting sustained repositioning that can extend over several weeks. The longest observed streak of 80 consecutive days corresponds to a single satellite (NORAD 59393) performing a sustained altitude reduction that caused the satellite to drift forward relative to the shell mean motion, resulting in continuous phasing detections throughout the movement duration.
Overall, across the major shells, most active satellites perform between one and five movements over the 174-day period.

\subsubsection{\textbf{ Movement Classification}} 
Understanding what drives satellite movements is central to characterizing real constellation dynamics. We classify each detected movement into one of three categories: collision avoidance, initial positioning, or other, by cross-referencing against CelesTrak SOCRATES \cite{celestrak_socrates} conjunction reports and satellite first-appearance data.

A movement is classified as \textit{collision avoidance} if the satellite appears in a conjunction report with a maximum collision probability of at least $3 \times 10^{-7}$ within a one-day window of the movement date --- aligned with SpaceX's reported operational threshold~\cite{aerospace_america_2025}. A movement is classified as \textit{initial positioning} if no conjunction match is found and it occurs within 15 days of the satellite's first appearance. This captures satellites that are actively adjusting their orbital slot upon entering the shell. Remaining movements are classified as \textit{other}.

\begin{figure}[h]
    \centering
    \includegraphics[width=0.8\columnwidth]{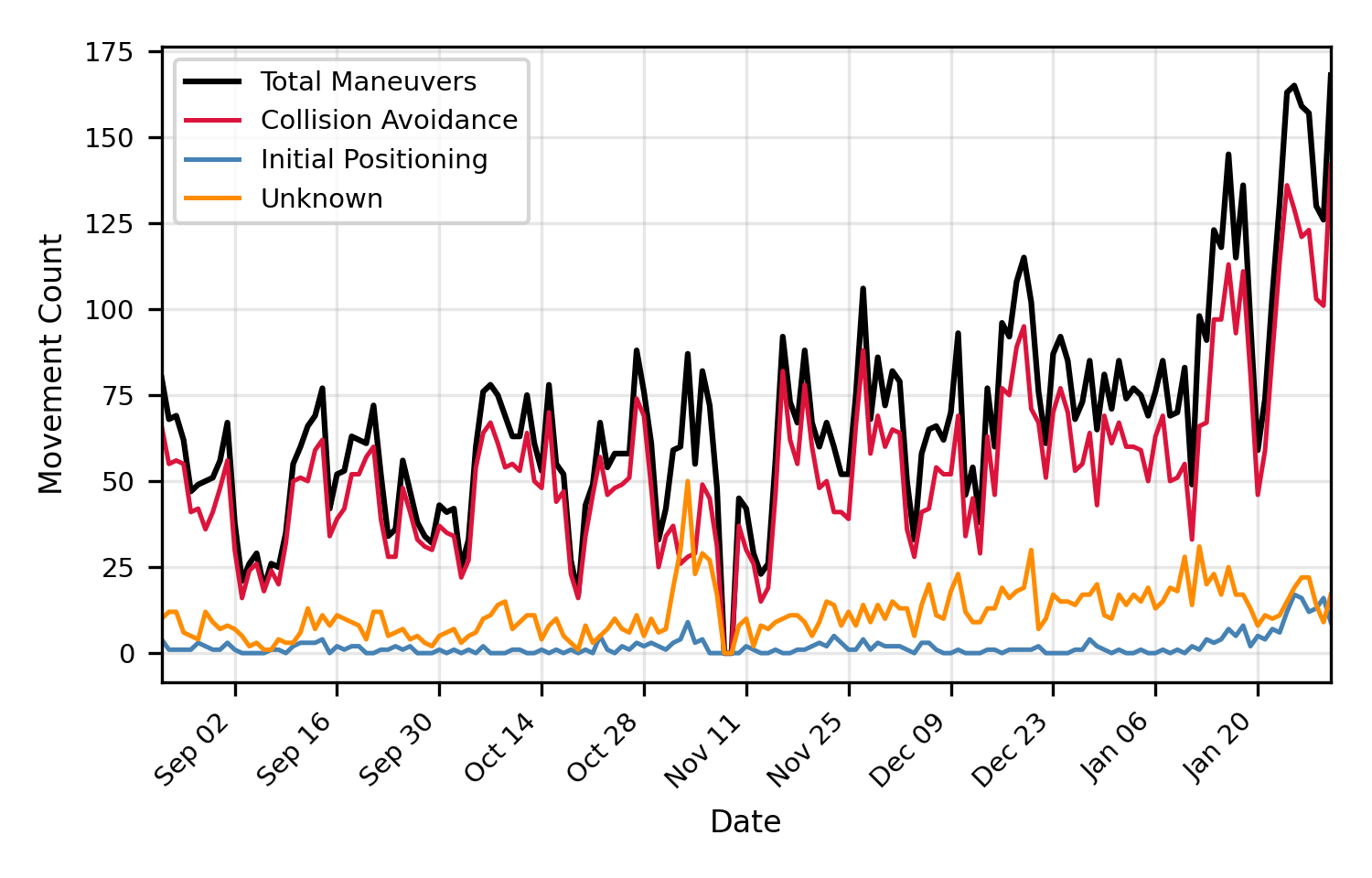}
    \caption{Movement classification over time for the 53.2°, 484 km shell.}
    \Description{Time series plot showing daily movement counts from September 2025 to January 2026, broken down into collision avoidance, initial positioning, and other categories.}
    \label{fig:classification_53_484}
\end{figure}

For the 53.2°, 484 km shell, collision avoidance dominates at 79.4\% (9,157 of 11,535 movements), with initial positioning at 2.9\% and other at 17.7\%. Figure~\ref{fig:classification_53_484} shows the daily breakdown; the collision avoidance line closely tracks total movement count throughout. This shows that the majority of detected movements are driven by conjunction activity.
Across all five major shells, collision avoidance accounts for 74.6\% of movements, averaging $\sim$197 per day. This is consistent with SpaceX's reported $\sim$275 collision avoidance maneuvers per day across the full constellation over a comparable period~\cite{spacex_50000_maneuvers}; our lower count is expected given detection is limited to five shells. Other movements constitute 24.3\%, and initial positioning 1.1\%.

Overall, our results show that the constellation is highly dynamic and that changes in altitude and orbital phasing are common across all major shells.

%% file: implications_network_design.tex
\section{Implications on Network Design}
\label{sec:implications_network_design}

Our analysis reveals that LEO shells are shaped by a range of real-world dynamics. Over time, launches, deorbits, and satellite movements change both shell population and the relative positions of satellites, while deployed shells often remain only partially populated and structurally asymmetric. These realities can affect all aspects of network operation, including topology design, routing, and network management. To demonstrate their practical impact, we use topology design as a case study and show that using the real deployed constellation can lead to markedly worse outcomes than assuming a fully deployed, perfectly regular one.

\subsection{Case Study: Impact on Network Topology}

A widely used topology design for LEO constellations is the \emph{$grid$} topology, in which each satellite connects to four inter-plane neighbors: two in the upward direction and two in the downward direction across adjacent orbital planes. This design assumes a symmetric constellation with uniformly spaced satellites and fully populated planes.

The real deployment of Starlink Shell--1 does not satisfy these assumptions. Because the shell is only partially deployed and the satellites are not evenly spaced, the ideal $grid$ cannot be applied directly. We therefore construct a \emph{$p$Grid} (partial-grid) topology that adapts the $grid$ design to the real shell by connecting each satellite to its nearest neighbors in adjacent planes. Due to partial deployment, some satellites do not find all four inter- and intra-plane neighbors.

\subsubsection{\textbf{Effect on topology metrics.}}
To understand the cost of this structural irregularity, we compare the average shortest path delay and hop count (between all satellite pairs) of $p$Grid against that of an idealized $grid$. We use the snapshot from 2022-01-01, when Shell--1 is close to its maximum deployment and is representative of a high-deployment regime. For comparison, the baseline $grid$ is evaluated on a synthetic fully deployed Shell--1 with evenly spaced satellites.

Compared to ideal $grid$, $p$Grid yields longer routes in both delay and hop count. The delay increases from 60.6~ms to 73.7~ms, a 17.8\% increase, while the average shortest-path hop count rises from 23.5 to 25.2 hops, a 6.7\% increase. This degradation arises because the irregular structure of the real shell limits inter-plane connectivity relative to the symmetric $grid$ design, forcing some routes to traverse longer network paths.

\begin{table}[ht]
\centering
\small
\begin{tabular}{l l c c}
\hline
\textbf{Src} & \textbf{Dest} & \textbf{$\Delta$ Delay (ms, \%)} & \textbf{$\Delta$ Hops} \\
\hline
Los Angeles & Berlin & 7.3 (17\%) & 2 \\
Rio de Janeiro & Cape Town & 13.5 (41\%) & 8 \\
Mumbai & Melbourne & 3.8 (10\%) & 3 \\
\hline
\end{tabular}
\caption{Examples of intercontinental city pairs whose shortest-path routes changed between January 1 and February 1, 2022 in Starlink Shell--1 under the $p$Grid topology.}
\label{tab:routing_changes_examples}
\end{table}

\subsubsection{\textbf{Temporal variability in routing.}}
Shell dynamics affect not only aggregate topology metrics but also the routes observed by end users over time. To evaluate this effect, we compare end-to-end routing between representative cities across six continents on two days one month apart, January 1 and February 1, 2022. For each city, we first identify visible satellites using a minimum elevation-angle threshold of $>25^\circ$. These satellite-ground links are then added to the topology graph, and shortest paths are computed over the $p$Grid topology derived from the satellite positions on each day.

Table~\ref{tab:routing_changes_examples} shows that routing characteristics can vary noticeably even for the same ground endpoints. For example, the Rio de Janeiro--Cape Town path changes by 13.5~ms, or 41\%, and requires eight additional hops. 
In an ideal symmetric $grid$, routes between fixed endpoints would remain largely stable over time. In the real Shell--1 deployment, however, partial deployment and asymmetry cause routes to encounter different parts of the topology at different times, leading to variations in both delay and hop count.

\subsubsection{\textbf{Link failures.}}

\begin{figure}[t]
\centering
\begin{minipage}{0.49\columnwidth}
\centering
\includegraphics[width=\linewidth]{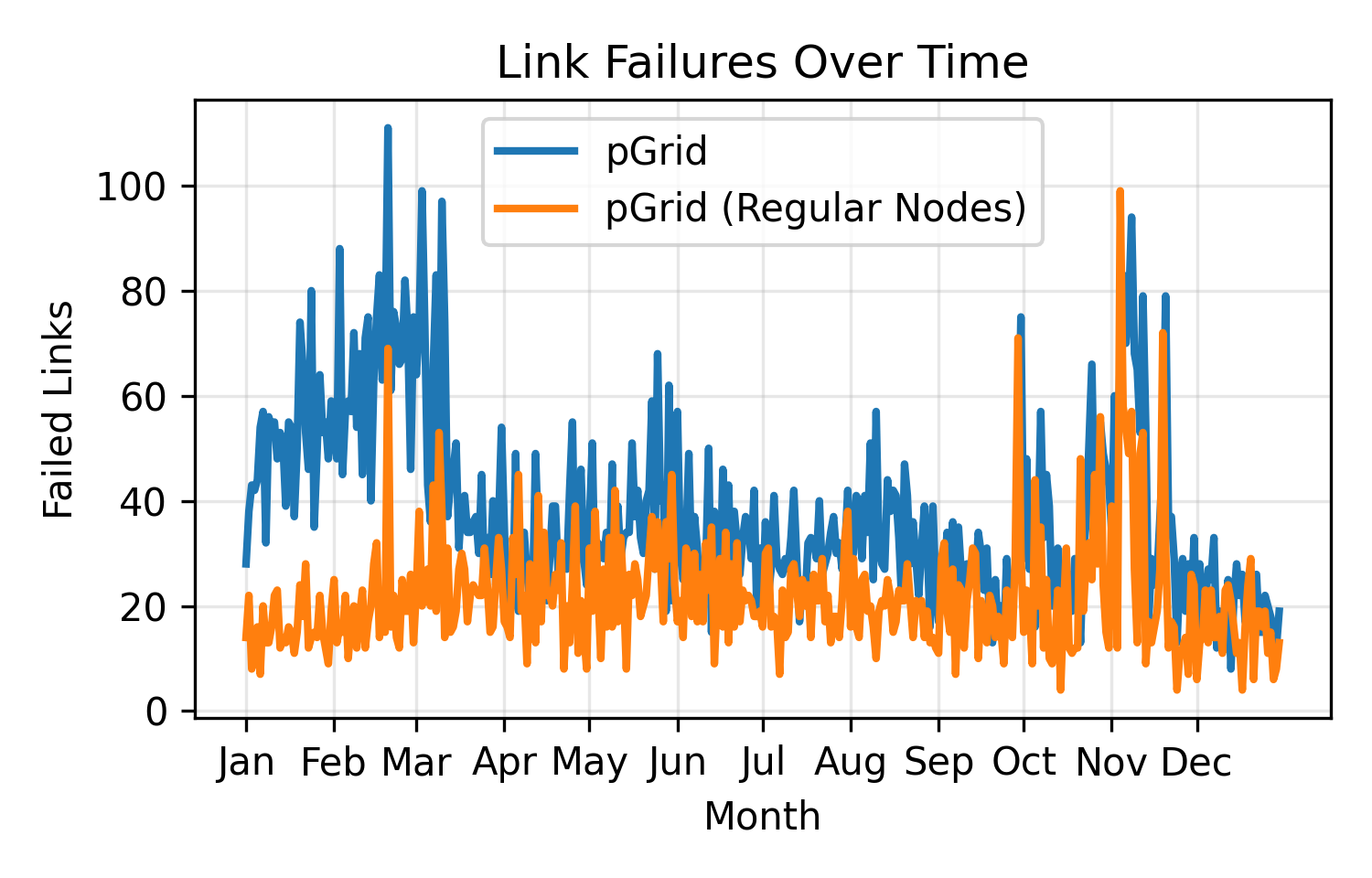}
\end{minipage}
\hfill
\begin{minipage}{0.49\columnwidth}
\centering
\includegraphics[width=\linewidth]{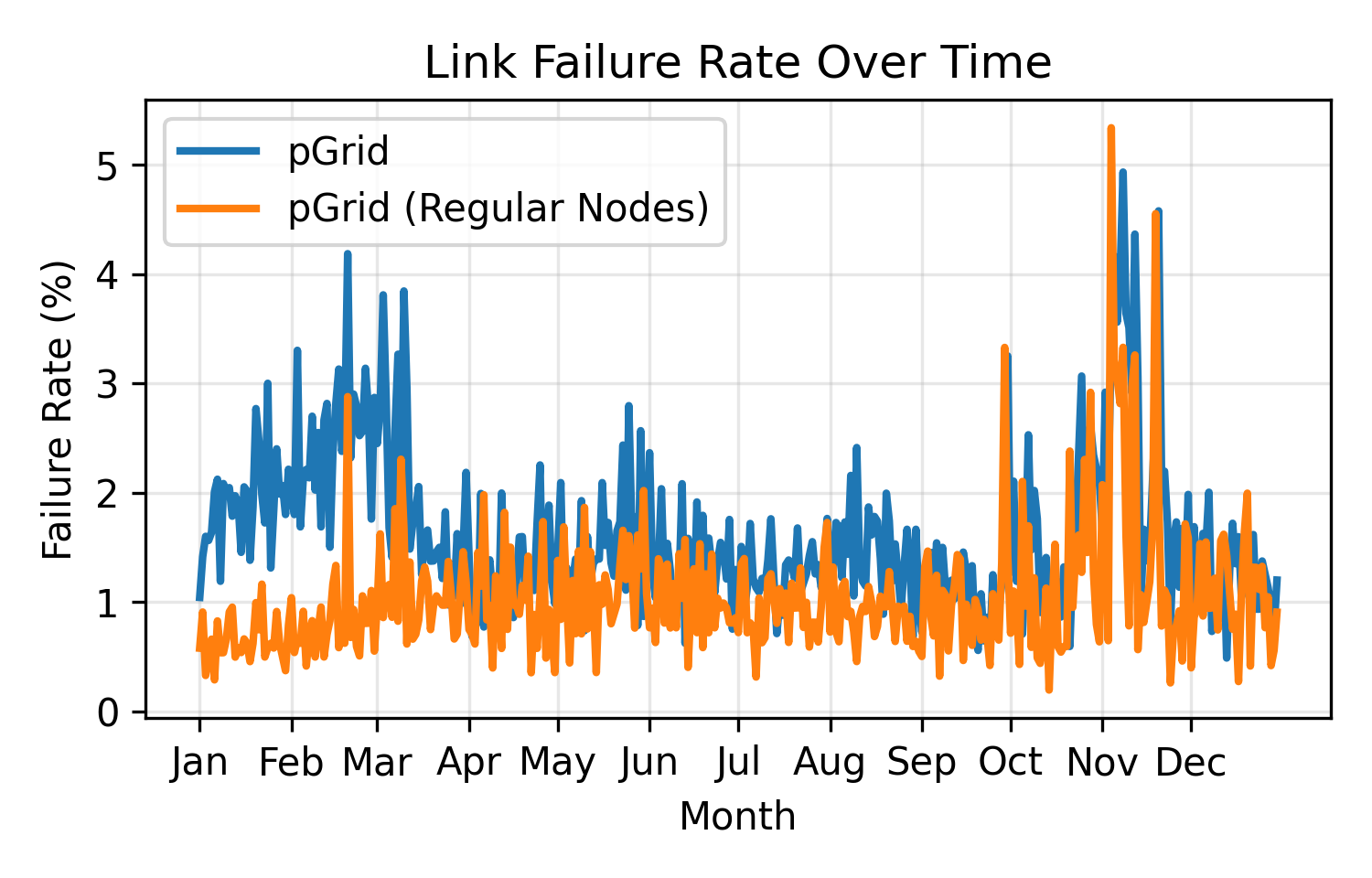}
\end{minipage}
\caption{Link failures observed in the $p$Grid topology over 2024 using the real deployment of Starlink Shell--1. The left figure shows the absolute number of daily link failures, while the right figure shows the corresponding failure rate.}
\label{fig:link_failures}
\end{figure}

Beyond changing route quality, shell dynamics also affect whether links themselves persist from one day to the next. Due to satellite movements, deorbits, replacements, and launches, the relative positions of satellites within a shell evolve over time. As a result, links that exist in the topology on one day may not persist on the following day. To capture this effect, we construct the $p$Grid topology for each day and propagate satellites to a common orbital phase so that their relative positions are approximately aligned across consecutive days. We then measure the links that fail across consecutive days.

Figure~\ref{fig:link_failures} shows this behavior for the real deployment of Starlink Shell--1 during 2024. When $p$Grid is built using all satellites, the topology experiences an average of $38$ link failures per day, corresponding to a failure rate of $1.63\%$. About $12\%$ of the failures are caused by satellites that are no longer present on the following day, while the rest ($88\%$) are due to satellite repositioning and movements that change relative neighbor relationships and alter the topology.

We then compare the $p$Grid topology constructed from all nodes with the $p$Grid topology constructed using only regular nodes. Using only the \emph{regular satellites} yields a substantially more stable topology. In this case (Figure~\ref{fig:link_failures}), the average number of link failures decreases to $22$ per day, and the failure rate drops to $1.04\%$.  $47\%$ of the failures are due to missing satellites, and the rest of the failures are due to satellite repositioning ($53\%$). These results show that the regularly placed satellites form a more stable backbone in the real deployed Shell--1, and that constructing the topology over this backbone can significantly reduce link churns.

Overall, these results show that deviations from an idealized shell structure can translate into noticeable changes in network behavior, highlighting the need for network design and evaluation to account for the realities of the deployed constellation. 

%% file: related_works.tex
\section{Related Work}
\label{sec:related_works}


A substantial body of work models LEO networks as idealized systems that are fully deployed, perfectly structured, and uniformly spaced when studying problems such as routing, topology design, and traffic engineering~\cite{motif,mcts,hetrogenous_topology,delyHandley,sate,routing_1,3_isl_grid,IPCGrid,hypatia,deepRL_routing}. As we show in this paper, however, the deployed constellation differs substantially from this idealized view.

Several studies have examined the performance and behavior of low Earth orbit (LEO) satellite networks, particularly Starlink, using measurement and modeling approaches. Measurement-based studies have focused on understanding the network performance experienced by users~\cite{starlink_performance}. Zhao et al.\ introduced \emph{LENS}, a dataset of Starlink performance measurements collected from 13 user terminals across seven points of presence on three continents, providing empirical insights into latency and throughput characteristics of the network~\cite{lens}. Izhikevich et al.\ proposed \emph{HitchHiking}, a methodology that infers Starlink’s network performance using Internet-exposed services~\cite{democratizeLeo}.  Li et al.\ evaluated the resilience of LEO networks by simulating failures of critical satellites, showing that the loss of key nodes can significantly degrade network performance and increase the workload on remaining satellites~\cite{resillienceLEO}. Zhang et al.\ analyzed how constellation design parameters, including orbit count, inclination, and satellite density, influence latency and hop count, identifying important thresholds for topology design~\cite{TopologyDesignParameters}.

 Oliveira et al.\ show that geomagnetic activity increases Starlink deborbit rates, linking space weather to satellite lifetime~\cite{Oliveira_2025}. Li et al.\ analyzed Starlink’s autonomous collision-avoidance mechanisms, highlighting how these maneuvers influence constellation stability and network operation~\cite{CollisionAvoidanceStarlink}. In contrast to prior work, we present the first comprehensive longitudinal study of Starlink's real deployment dynamics, examining in-orbit shell structure, satellite movements, satellite lifecycles, and their implications for network design using six years of observational data from 2019 to 2025.

%% file: conclusion.tex
\section{Conclusion}
\label{sec:conclusion}

In this paper, we present a comprehensive longitudinal study of Starlink, the de facto reference constellation for LEO networking research, using six years of satellite observations. Our analysis shows that the deployed constellation is far more dynamic and irregular than this idealized view suggests. We find that shells evolve over time, satellite lifetimes vary substantially, and daily satellite movements alter relative in-orbit phasing. We also uncover a mixed in-orbit structure in which regularly placed satellites coexist with non-regular satellites, including twins and triads, which likely serve as backup nodes. To demonstrate the networking implications of these dynamics, we conduct a case study on topology design under real deployment conditions. We show that deployment irregularity, node turnover, and satellite movement materially affect delay, routing behavior, and link churn. These findings highlight the need for future LEO network designs to move beyond idealized constellation models and explicitly account for the realities of deployed systems.

%% file: appendix.tex
\appendix
\section{Supplementary Figures}

This section contains any supplementary figures linked to the analysis.
\begin{figure*}[h!]
\centering
\includegraphics[width=\textwidth]{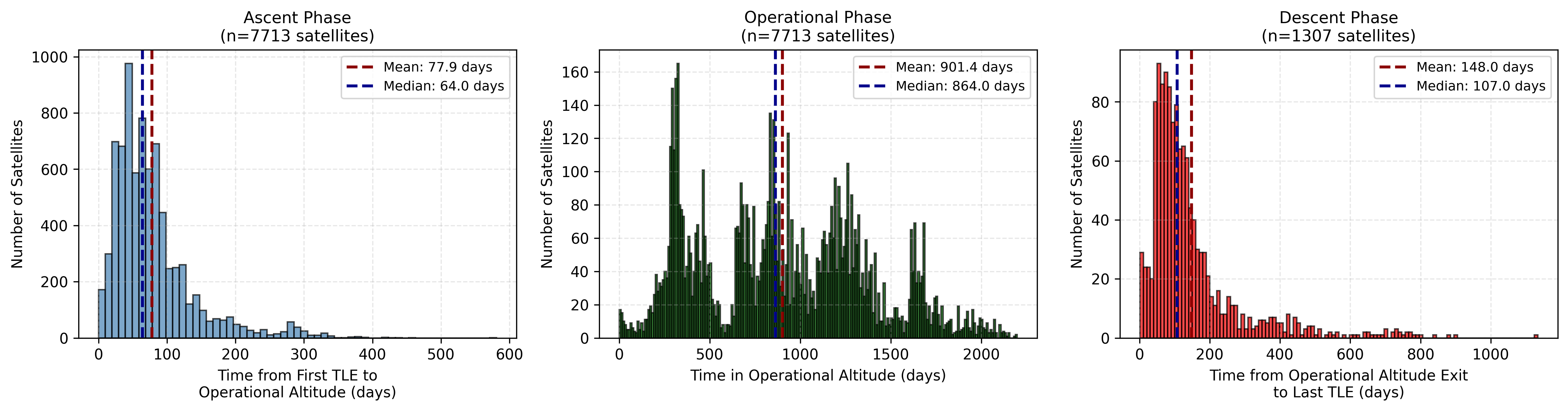}
\caption{Distribution of satellite lifecycle phases}
\label{fig:lifecycle}
\end{figure*}

\begin{figure}[h]
    \centering
    \includegraphics[width=0.7\linewidth]{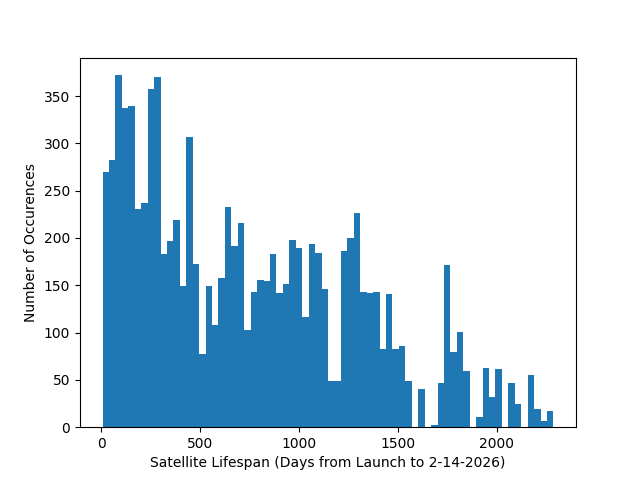}
    \caption{Age of  Satellites}
    \label{Lifespan of active satellites}
\end{figure}

%% file: references.bib
@Article{deepRL_routing,
AUTHOR = {Han, Chi and Xiong, Wei and Yu, Ronghuan},
TITLE = {Deep Reinforcement Learning-Based Multipath Routing for LEO Megaconstellation Networks},
JOURNAL = {Electronics},
VOLUME = {13},
YEAR = {2024},
NUMBER = {15},
ARTICLE-NUMBER = {3054},
URL = {https://www.mdpi.com/2079-9292/13/15/3054},
ISSN = {2079-9292},
DOI = {10.3390/electronics13153054}
}

@inproceedings{hypatia,
author = {Kassing, Simon and Bhattacherjee, Debopam and \'{A}guas, Andr\'{e} Baptista and Saethre, Jens Eirik and Singla, Ankit},
title = {Exploring the "Internet from space" with Hypatia},
year = {2020},
isbn = {9781450381383},
publisher = {Association for Computing Machinery},
address = {New York, NY, USA},
url = {https://doi.org/10.1145/3419394.3423635},
doi = {10.1145/3419394.3423635},
booktitle = {Proceedings of the ACM Internet Measurement Conference},
pages = {214–229},
numpages = {16},
location = {Virtual Event, USA},
series = {IMC '20}
}

@ARTICLE{3_isl_grid,
  author={Chen, Quan and Yang, Lei and Zhao, Yong and Wang, Yi and Zhou, Haibo and Chen, Xiaoqian},
  journal={IEEE Transactions on Aerospace and Electronic Systems}, 
  title={3-ISL Topology: Routing Properties and Performance in LEO Megaconstellation Networks}, 
  year={2025},
  volume={61},
  number={2},
  pages={4961-4972},
  doi={10.1109/TAES.2024.3512535}}

@inproceedings{starlink_performance,
author = {Michel, Fran\c{c}ois and Trevisan, Martino and Giordano, Danilo and Bonaventure, Olivier},
title = {A first look at starlink performance},
year = {2022},
isbn = {9781450392594},
publisher = {Association for Computing Machinery},
address = {New York, NY, USA},
url = {https://doi.org/10.1145/3517745.3561416},
doi = {10.1145/3517745.3561416},
booktitle = {Proceedings of the 22nd ACM Internet Measurement Conference},
pages = {130–136},
numpages = {7},
location = {Nice, France},
series = {IMC '22}
}

@INPROCEEDINGS{routing_1,
  author={Henderson, T.R. and Katz, R.H.},
  booktitle={Globecom '00 - IEEE. Global Telecommunications Conference. Conference Record (Cat. No.00CH37137)}, 
  title={On distributed, geographic-based packet routing for LEO satellite networks}, 
  year={2000},
  volume={2},
  number={},
  pages={1119-1123 vol.2},
  doi={10.1109/GLOCOM.2000.891311}}

@inproceedings{sate,
author = {Wu, Hao and Han, Yizhan and Rajpal, Mohit and Zhang, Qizhen and Wang, Jingxian},
title = {SaTE: Low-Latency Traffic Engineering for Satellite Networks},
year = {2025},
isbn = {9798400715242},
publisher = {Association for Computing Machinery},
address = {New York, NY, USA},
url = {https://doi.org/10.1145/3718958.3750524},
doi = {10.1145/3718958.3750524},
booktitle = {Proceedings of the ACM SIGCOMM 2025 Conference},
pages = {896–916},
numpages = {21},
location = {S\~{a}o Francisco Convent, Coimbra, Portugal},
series = {SIGCOMM '25}
}

@misc{orbital_parameters, 
  title        = {Orbital and Technical Parameters | European GNSS Service Centre
},
url={https://www.gsc-europa.eu/system-service-status/orbital-and-technical-parameters}, journal={Orbital and Technical Parameters | European GNSS Service Centre (GSC)}}

@misc{hetrogenous_topology, title={Efficient topology design for LEO mega-constellation using topological structure units with heterogeneous isls}, url={https://pmc.ncbi.nlm.nih.gov/articles/PMC12473503/#abstract1}, journal={Sensors (Basel, Switzerland)}, publisher={U.S. National Library of Medicine}, author={Zhang, Wei and Wu, Tao and Yan, Xucun and Li, Guixin and Ma, Hongbin}, year={2025}, month={Sep}}

@inproceedings{motif,
author = {Bhattacherjee, Debopam and Singla, Ankit},
title = {Network topology design at 27,000 km/hour},
year = {2019},
isbn = {9781450369985},
publisher = {Association for Computing Machinery},
address = {New York, NY, USA},
url = {https://doi.org/10.1145/3359989.3365407},
doi = {10.1145/3359989.3365407},
booktitle = {Proceedings of the 15th International Conference on Emerging Networking Experiments And Technologies},
pages = {341–354},
numpages = {14},
location = {Orlando, Florida},
series = {CoNEXT '19}
}

@inproceedings{delyHandley,
author = {Handley, Mark},
title = {Delay is Not an Option: Low Latency Routing in Space},
year = {2018},
isbn = {9781450361200},
publisher = {Association for Computing Machinery},
address = {New York, NY, USA},
url = {https://doi.org/10.1145/3286062.3286075},
doi = {10.1145/3286062.3286075},
booktitle = {Proceedings of the 17th ACM Workshop on Hot Topics in Networks},
pages = {85–91},
numpages = {7},
location = {Redmond, WA, USA},
series = {HotNets '18}
}

@article{IPCGrid,
author = {Chen, Quan and Giambene, Giovanni and Yang, Lei and Fan, Chengguang and Chen, Xiaoqian},
year = {2021},
month = {03},
pages = {2743 - 2755},
title = {Analysis of Inter-Satellite Link Paths for LEO Mega-Constellation Networks},
volume = {70},
journal = {IEEE Transactions on Vehicular Technology},
doi = {10.1109/TVT.2021.3058126}
}

@ARTICLE{mcts,
author={Hu, Han and Lyu, Yifeng and Song, Kaifeng and Fan, Rongfei and Zhan, Cheng and Yang, Jian},
journal={ IEEE Transactions on Mobile Computing },
title={{ An Efficient Two-Stage Networking Topology Design for Mega-Constellation of Low Earth Orbit Satellites }},
year={2025},
volume={24},
number={07},
ISSN={1558-0660},
pages={6333-6347},
doi={10.1109/TMC.2025.3540671},
url = {https://doi.ieeecomputersociety.org/10.1109/TMC.2025.3540671},
publisher={IEEE Computer Society},
address={Los Alamitos, CA, USA},
month=jul}

@misc{spacetrack_usspacecom,
  title        = {Space-Track.org: Public Access to Satellite Data and Space Surveillance Information},
  howpublished = {\url{https://www.space-track.org/}},
  note         = { Accessed: 2025-12-01}
}

@inproceedings{lens,
author = {Zhao, Jinwei and Pan, Jianping},
title = {LENS: A LEO Satellite Network Measurement Dataset},
year = {2024},
isbn = {9798400704123},
publisher = {Association for Computing Machinery},
address = {New York, NY, USA},
url = {https://doi.org/10.1145/3625468.3652170},
doi = {10.1145/3625468.3652170},
booktitle = {Proceedings of the 15th ACM Multimedia Systems Conference},
pages = {278–284},
numpages = {7},
location = {Bari, Italy},
series = {MMSys '24}
}

@article{democratizeLeo,
author = {Izhikevich, Liz and Tran, Manda and Izhikevich, Katherine and Akiwate, Gautam and Durumeric, Zakir},
title = {Democratizing LEO Satellite Network Measurement},
year = {2024},
issue_date = {March 2024},
publisher = {Association for Computing Machinery},
address = {New York, NY, USA},
volume = {8},
number = {1},
url = {https://doi.org/10.1145/3639039},
doi = {10.1145/3639039},
journal = {Proc. ACM Meas. Anal. Comput. Syst.},
month = feb,
articleno = {13},
numpages = {26}
}

@inproceedings{resillienceLEO,
author = {Li, Zhuoyuan and Zhang, Wenyi Morty and Chen, Wenhao and Hu, Yiyan and Lu, Weyl},
title = {LEO Satellite Network Resilience Analysis: A Focus on Critical Satellites},
year = {2024},
isbn = {9798400712807},
publisher = {Association for Computing Machinery},
address = {New York, NY, USA},
url = {https://doi.org/10.1145/3697253.3697267},
doi = {10.1145/3697253.3697267},
booktitle = {Proceedings of the 2nd International Workshop on LEO Networking and Communication},
pages = {13–18},
numpages = {6},
location = {Washington, DC, USA},
series = {LEO-NET '24}
}

@inproceedings{TopologyDesignParameters,
author = {Zhang, Wenyi and Xu, Zihan and Jyothi, Sangeetha Abdu},
title = {An In-Depth Investigation of LEO Satellite Topology Design Parameters},
year = {2024},
isbn = {9798400712807},
publisher = {Association for Computing Machinery},
address = {New York, NY, USA},
url = {https://doi.org/10.1145/3697253.3697263},
doi = {10.1145/3697253.3697263},
booktitle = {Proceedings of the 2nd International Workshop on LEO Networking and Communication},
pages = {1–6},
numpages = {6},
location = {Washington, DC, USA},
series = {LEO-NET '24}
}

@article{Oliveira_2025,
   title={Tracking reentries of Starlink satellites during the rising phase of solar cycle 25},
   volume={12},
   ISSN={2296-987X},
   url={http://dx.doi.org/10.3389/fspas.2025.1572313},
   DOI={10.3389/fspas.2025.1572313},
   journal={Frontiers in Astronomy and Space Sciences},
   publisher={Frontiers Media SA},
   author={Oliveira, Denny M. and Zesta, Eftyhia and Garcia-Sage, Katherine},
   year={2025},
   month=jun }

@inproceedings{CollisionAvoidanceStarlink,
author = {Li, Yuanjie and Li, Hewu and Liu, Wei and Liu, Lixin and Zhao, Wei and Chen, Yimei and Wu, Jianping and Wu, Qian and Liu, Jun and Lai, Zeqi and Qiu, Han},
title = {A Networking Perspective on Starlink's Self-Driving LEO Mega-Constellation},
year = {2023},
isbn = {9781450399906},
publisher = {Association for Computing Machinery},
address = {New York, NY, USA},
url = {https://doi.org/10.1145/3570361.3592519},
doi = {10.1145/3570361.3592519},
booktitle = {Proceedings of the 29th Annual International Conference on Mobile Computing and Networking},
articleno = {17},
numpages = {16},
location = {Madrid, Spain},
series = {ACM MobiCom '23}
}

@misc{spacex_50000_maneuvers,
  author       = {Pultarova, Tereza},
  title        = {{SpaceX Starlink} satellites made 50,000 collision-avoidance maneuvers in the past 6 months},
  year         = {2024},
  month        = {July},
  howpublished = {Space.com},
  url          = {https://www.space.com/spacex-starlink-50000-collision-avoidance-maneuvers-space-safety},
}

@misc{aerospace_america_2025,
  author       = {O'Callaghan, Jonathan},
  title        = {Heavy Traffic Ahead},
  year         = {2025},
  month        = {October},
  howpublished = {Aerospace America},
  url          = {https://aerospaceamerica.aiaa.org/features/heavy-traffic-ahead/},
}

@misc{fcc_gen2_partial_grant,
  author       = {{Federal Communications Commission}},
  title        = {Authorization: SpaceX Second-Generation {Starlink} Constellation, {FCC} 22-91},
  year         = {2022},
  month        = {December},
  howpublished = {Federal Communications Commission},
  url          = {https://docs.fcc.gov/public/attachments/FCC-22-91A1.pdf},
}

@misc{fcc_low_alt,
  author       = {{Federal Communications Commission}},
  title        = {Authorization: {SpaceX} {Gen2} {Starlink} Constellation Modification, {DA} 26-36A1},
  year         = {2026},
  month        = {November},
  howpublished = {Federal Communications Commission},
  url          = {https://docs.fcc.gov/public/attachments/DA-26-36A1.pdf},
}

@article{foust_2026_lowering,
  author       = {Foust, Jeff},
  title        = {{SpaceX} to lower orbits of some {Starlink} satellites},
  year         = {2026},
  month        = {January},
  journal      = {SpaceNews},
  url          = {https://spacenews.com/spacex-to-lower-orbits-of-some-starlink-satellites/},
}

@misc{gunter_starlink_v15,
  author       = {Krebs, Gunter Dirk},
  title        = {Starlink Block v1.5},
  howpublished = {Gunter's Space Page},
  url          = {https://space.skyrocket.de/doc_sdat/starlink-v1-5.htm},
  note         = {Accessed: 2026}
}

@article{kaplan1958nonparametric,
  title={Nonparametric estimation from incomplete observations},
  author={Kaplan and Meier Paul},
  journal={Journal of the American Statistical Association},
  volume={53},
  number={282},
  year={1958}
}

@inproceedings{10.5555/3001460.3001507,
author = {Ester, Martin and Kriegel, Hans-Peter and Sander, J\"{o}rg and Xu, Xiaowei},
title = {A density-based algorithm for discovering clusters in large spatial databases with noise},
year = {1996},
publisher = {AAAI Press},
booktitle = {Proceedings of the Second International Conference on Knowledge Discovery and Data Mining},
pages = {226–231},
numpages = {6},
location = {Portland, Oregon},
series = {KDD'96}
}

@techreport{fcc2021starlink,
  title={{FCC} 21-48: {SpaceX} Third Modification Order},
  author={{Federal Communications Commission}},
  year={2021},
  institution={Federal Communications Commission},
  url={https://docs.fcc.gov/public/attachments/fcc-21-48a1.pdf}
}

@techreport{fcc_eratum23,
  author       = {{Federal Communications Commission}},
  title        = {{Request for Orbital Deployment and Operating
Authority for the SpaceX Gen2 NGSO Satellite
System}},
  institution  = {{Federal Communications Commission}},
  year         = {2023},
  month        = feb,
  url          = {https://docs.fcc.gov/public/attachments/DOC-390793A1.pdf},
}

@misc{celestrak_socrates,
  author       = {Kelso, T.S.},
  title        = {{SOCRATES}: Satellite Orbital Conjunction Reports Assessing Threatening Encounters in Space},
  howpublished = {\url{https://celestrak.org/SOCRATES/}},
  note         = {Accessed: 2026},
  organization = {CelesTrak}
}
